\documentclass[twocolumn,showpacs,preprintnumbers,amsmath,amssymb]{revtex4}

\usepackage{graphicx}
\usepackage{dcolumn}
\usepackage{bm}

\usepackage{amssymb}
\usepackage{amsmath}

\begin{document}


\title{Holographic thermodynamic relation for dissipative and non-dissipative universes \\ in a flat FLRW cosmology}

\author{Nobuyoshi {\sc Komatsu}}  \altaffiliation{E-mail: komatsu@se.kanazawa-u.ac.jp} 
\affiliation{Department of Mechanical Systems Engineering, Kanazawa University, Kakuma-machi, Kanazawa, Ishikawa 920-1192, Japan}

\date{\today}

\begin{abstract}

Horizon thermodynamics and cosmological equations in standard cosmology provide a holographic-like connection between thermodynamic quantities on a cosmological horizon and in the bulk.
It is expected that this connection can be modified as a holographic-like thermodynamic relation for dissipative and non-dissipative universes whose Hubble volume $V$ varies with time $t$.
To clarify such a modified thermodynamic relation, the present study applies a general formulation for cosmological equations in a flat Friedmann--Lema\^{i}tre--Robertson--Walker (FLRW) universe to the first law of thermodynamics, using the Bekenstein--Hawking entropy $S_{\rm{BH}}$ and a dynamical Kodama--Hayward temperature $T_{\rm{KH}}$.
For the general formulation, both an effective pressure $p_{e}$ of cosmological fluids for dissipative universes (e.g., bulk viscous cosmology) and an extra driving term $f_{\Lambda}(t)$ for non-dissipative universes (e.g., time-varying $\Lambda (t)$ cosmology) are phenomenologically assumed.
A modified thermodynamic relation is derived by applying the general formulation to the first law,
which includes both $p_{e}$ and an additional time-derivative term $\dot{f}_{\Lambda}(t)$, related to a non-zero term of the general continuity equation.
When $f_{\Lambda}(t)$ is constant, the modified thermodynamic relation is equivalent to the formulation of the first law in standard cosmology.
One side of this modified relation describes thermodynamic quantities in the bulk and can be divided into two time-derivative terms, namely $\dot{\rho}$ and $\dot{V}$ terms, where $\rho$ is the mass density of cosmological fluids.
Using the Gibbons--Hawking temperature $T_{\rm{GH}}$, the other side of this relation, $T_{\rm{KH}}  \dot{S}_{\rm{BH}}$, can be formulated as the sum of $T_{\rm{GH}}  \dot{S}_{\rm{BH}}$ and $[(T_{\rm{KH}}/T_{\rm{GH}}) -1] T_{\rm{GH}}  \dot{S}_{\rm{BH}}$, which are equivalent to the $\dot{\rho}$ and $\dot{V}$ terms, respectively,
with the magnitude of the $\dot{V}$ term being proportional to the square of the $\dot{\rho}$ term.
In addition, the modified thermodynamic relation for constant $f_{\Lambda}(t)$ is examined by applying the equipartition law of energy on the horizon.
This modified thermodynamic relation reduces to a kind of extended holographic-like connection when a constant $T_{\rm{KH}}$ universe (whose Hubble volume varies with time) is considered.
The evolution of thermodynamic quantities is also discussed, using a constant $T_{\rm{KH}}$ model, extending a previous analysis [N. Komatsu, Phys. Rev. D \textbf{108}, 083515 (2023)].

\end{abstract}


\maketitle

\section{Introduction} 
\label{Introduction}

To explain the accelerated expansion of the late Universe \cite{PERL1998_Riess1998,Planck2018,Hubble2017}, astrophysicists have proposed various cosmological models, 
e.g., lambda cold dark matter ($\Lambda$CDM) models, time-varying $\Lambda (t)$ cosmology \cite{FreeseOverduin,Nojiri2006etc,Valent2015Sola2019,Sola_2009-2022}, creation of CDM (CCDM) models \cite{Prigogine_1988-1989,Lima1992-1996,LimaOthers2023,Freaza2002Cardenas2020}, 
bulk viscous cosmology \cite{BarrowLima,BrevikNojiri,EPJC2022}, and thermodynamic scenarios such as entropic cosmology \cite{EassonCai,Basilakos1,Koma45,Koma6,Koma7,Koma8,Koma9,Neto2022,Gohar2024}.
These studies imply that our Universe should finally approach a de Sitter universe whose horizon is considered to be in thermal equilibrium.
The thermodynamics of the universe has been examined from various perspectives, 
e.g., the first law of thermodynamics \cite{Cai2005,Cai2011,Dynamical-T-2007,Dynamical-T-20092014,Sheykhi1,Sheykhi2Karami,Santos2022,Sheykhia2018,ApparentHorizon2022,Cai2007,Cai2007B,Cai2008,Sanchez2023,Nojiri2024,Odintsov2023ab,Odintsov2024,Mohammadi2023,Odintsov2024B}, 
the second law of thermodynamics \cite{Easther1-Egan1,Pavon2013Mimoso2013,Bamba2018Pavon2019,deSitter_entropy,Saridakis2019,Saridakis2021,Sharif2024}, and the holographic equipartition law related to the emergence of cosmic space 
\cite{Padma2010,Verlinde1,HDE,Padmanabhan2004,ShuGong2011,Koma14,Koma15,Koma16,Koma17,Koma19,Koma20,Padma2012AB,Cai2012,Hashemi,Moradpour,Wang,Koma10,Koma11,Koma12,Koma18,Krishna20172019,Mathew2022,Chen2022,Luciano,Mathew2023,Mathew2023b,Pad2017,Tu2018,Tu2019,Chen2024}.

In the thermodynamic scenarios, black hole thermodynamics \cite{Hawking1Bekenstein1} is applied to a cosmological horizon and the information of the bulk is assumed to be stored on the horizon, based on the holographic principle \cite{Hooft-Bousso}.
In particular, the first law of thermodynamics has been examined from a holographic viewpoint \cite{Cai2005,Cai2011,Dynamical-T-2007,Dynamical-T-20092014,Sheykhi1,Sheykhi2Karami,Santos2022,Sheykhia2018,ApparentHorizon2022,Cai2007,Cai2007B,Cai2008,Sanchez2023,Nojiri2024,Odintsov2023ab,Odintsov2024,Mohammadi2023,Odintsov2024B}.
In these works (excepting Refs.\
 \cite{Santos2022,Mohammadi2023}), the Friedmann equation is derived from the first law, using the continuity equation whose right-hand side is considered to be zero.
Of course, it is well known that  the continuity equation can be non-zero in cosmological models for both dissipative universes (e.g., bulk viscous models) and non-dissipative universes (e.g., $\Lambda(t)$CDM models) \cite{Koma9}.
In fact, a general formulation for the cosmological equations of the two types of universes has been examined in previous works \cite{Koma9,Koma16}.
We expect that a holographic thermodynamic relation for dissipative and non-dissipative universes can be derived by applying the general formulation to the first law of thermodynamics.

In addition, Padmanabhan \cite{Pad2017} has derived an energy-balance relation using the equipartition law of energy on the horizon.
A similar holographic-like connection relation $E_{\rm{bulk}}=F_{H}$ has been recently examined \cite{Koma18}, where $E_{\rm{bulk}}$ is an energy in the bulk and $F_{H}$ is the Helmholtz free energy on the horizon.
(For details on the holographic-like connection, see Appendix\ \ref{Holographic-like connection} and Ref.\ \cite{Koma18}.)
In these works, de Sitter universes are originally considered and, therefore, the Hubble parameter, the Hubble volume, and the Gibbons--Hawking temperature $T_{\rm{GH}}$  \cite{GibbonsHawking1977} are constant.
In a de Sitter universe, $T_{\rm{GH}}$ is equivalent to the dynamical Kodama--Hayward temperature $T_{\rm{KH}}$ \cite{Dynamical-T-2007,Dynamical-T-20092014}, based on the works of Hayward \textit{et al}. \cite{Dynamical-T-1998,Dynamical-T-2008}.
Of course, the horizons of universes are generally considered to be dynamic, unlike for de Sitter universes.
Accordingly, a dynamical temperature should be appropriate for discussing the thermodynamics on a dynamic horizon \cite{Koma19}. 

The first law of thermodynamics has been recently examined using the dynamical Kodama--Hayward temperature $T_{\rm{KH}}$.
(For the first law, see the previous works of Akbar and Cai \cite{Cai2007,Cai2007B} and Cai \textit{et al.} \cite{Cai2008} and recent works of S\'{a}nchez and Quevedo \cite{Sanchez2023}, Nojiri \textit{et al.} \cite{Nojiri2024}, and Odintsov \textit{et al.} \cite{Odintsov2023ab,Odintsov2024,Odintsov2024B}.)
The first law should lead to an extended holographic-like connection, namely a modified thermodynamic relation for dissipative and non-dissipative universes whose Hubble volume varies with time.
We expect that the thermodynamic relation between the horizon and the bulk can reduce to a simple relation similar to the holographic-like connection by applying the equipartition law of energy on the horizon.

However, the modified thermodynamic relation has not yet been discussed from those viewpoints.
An understanding of the thermodynamic relation for dissipative and non-dissipative universes should provide new insights into the thermodynamics on the horizon and cosmological equations in the bulk.
In this context, we examine the thermodynamic relation by applying a general formulation for cosmological equations to the first law of thermodynamics.

The remainder of the present article is organized as follows.
In Sec.\ \ref{Cosmological equations}, a general formulation for cosmological equations is reviewed.
In addition, an associated entropy and an approximate temperature on a cosmological horizon are introduced.
In Sec.\ \ref{The first law}, the general formulation is applied to the first law of thermodynamics to derive a modified thermodynamic relation.
In Sec.\ \ref{A modified thermodynamic relation}, the modified thermodynamic relation is discussed under a specific condition.
In Sec.\ \ref{drho and dV terms}, the left-hand side of the modified thermodynamic relation, corresponding to thermodynamic quantities in the bulk, is examined.
In Sec.\ \ref{Equipartition law}, the equipartition law of energy on the horizon is applied to the right-hand side of the relation, namely thermodynamic quantities on the horizon.
In Sec.\ \ref{A constant TKH model}, typical evolutions of the thermodynamic quantities in the relation are observed using cosmological models.
Finally, in Sec.\ \ref{Conclusions}, the conclusions of the study are presented.

In this paper, a homogeneous, isotropic, and spatially flat universe, namely a flat Friedmann--Lema\^{i}tre--Robertson--Walker (FLRW) universe, is considered.
Therefore, the apparent horizon of the universe is equivalent to the Hubble horizon.
Also, an expanding universe is assumed from observations \cite{Hubble2017}.
Inflation of the early universe and density perturbations related to structure formations are not discussed.

\section{Cosmological equations, horizon entropy, and horizon temperature} 
\label{Cosmological equations}

In the present study, a general formulation for cosmological equations is applied to the first law of thermodynamics.
For this, Sec.\ \ref{General formulation} reviews the general formulation, while
in Sec.\ \ref{Entropy and temperature}, the Bekenstein--Hawking entropy, the Gibbons--Hawking temperature, and a dynamical Kodama--Hayward temperature on the Hubble horizon are introduced.

\subsection{General formulation for cosmological equations in a flat FLRW universe} 
\label{General formulation}

We introduce a general formulation for cosmological equations in dissipative and non-dissipative universes, using the scale factor $a(t)$ at time $t$, based on previous works \cite{Koma6,Koma9,Koma14,Koma15,Koma16}.
The general Friedmann, acceleration, and continuity equations are written as 
\begin{equation}
 H(t)^2      =  \frac{ 8\pi G }{ 3 } \rho (t)    + f_{\Lambda}(t)            ,                                                 
\label{eq:General_FRW01} 
\end{equation} 
\begin{align}
  \frac{ \ddot{a}(t) }{ a(t) }   
                                          &= \dot{H}(t) + H(t)^{2}           \notag    \\ 
                                          &= -  \frac{ 4\pi G }{ 3 }  \left ( \rho (t) +  \frac{3 p(t)}{c^2} \right )                   +   f_{\Lambda}(t)    +  h_{\textrm{B}}(t)  , 
\label{eq:General_FRW02}
\end{align}
\begin{equation}
       \dot{\rho} + 3  H \left ( \rho (t) +  \frac{p(t)}{c^2} \right )       =    -  \frac{3}{8 \pi G }   \dot{f}_{\Lambda}(t)      +    \frac{3 }{4 \pi G}     H h_{\textrm{B}}(t)              , 
\label{eq:drho_General}
\end{equation}
with the Hubble parameter $H(t)$ defined as 
\begin{equation}
   H(t) \equiv   \frac{ da/dt }{a(t)} =   \frac{ \dot{a}(t) } {a(t)}  ,
\label{eq:Hubble}
\end{equation}
where $G$, $c$, $\rho(t)$, and $p(t)$ are the gravitational constant, the speed of light, the mass density of cosmological fluids, and the pressure of cosmological fluids, respectively.
Two extra driving terms, $f_{\Lambda}(t)$ and $h_{\textrm{B}}(t)$, are phenomenologically assumed \cite{Koma14}.
Specifically, $f_{\Lambda}(t)$ is used for a $\Lambda (t)$ model, similar to $\Lambda(t)$CDM models, whereas $h_{\textrm{B}}(t)$ is used for a BV (bulk-viscous-cosmology-like) model, similar to bulk viscous models and CCDM models \cite{Koma14,Koma16}.
That is, $f_{\Lambda}(t)$ is used for non-dissipative universes and $h_{\textrm{B}}(t)$ is used for dissipative universes.
In this study, $f_{\Lambda}(t)$ and $h_{\textrm{B}}(t)$ are considered simultaneously.
Only two of the three equations (the Friedmann, acceleration, and continuity equations) are independent \cite{Ryden1}. 
Therefore, the general continuity equation given by Eq. (\ref{eq:drho_General}) can be derived from Eqs.\ (\ref{eq:General_FRW01}) and (\ref{eq:General_FRW02}).
In addition, subtracting Eq.\ (\ref{eq:General_FRW01}) from Eq.\ (\ref{eq:General_FRW02}) yields 
\begin{equation}
    \dot{H} = - 4\pi G  \left ( \rho (t) +  \frac{p(t)}{c^2} \right )  + h_{\textrm{B}}(t)   .
\label{eq:dotH}
\end{equation}
These equations are used in Sec.\ \ref{The first law}.

Equation\ (\ref{eq:drho_General}) indicates that the right-hand side of the general continuity equation is non-zero.
A similar non-zero term appears in other cosmological models, such as energy exchange cosmology \cite{Barrow22,Wang0102,Dynamical20052013} and the bulk viscous and CCDM models \cite{Prigogine_1988-1989,Lima1992-1996,LimaOthers2023,Freaza2002Cardenas2020,BarrowLima,BrevikNojiri,EPJC2022}, as discussed in Refs.\ \cite{Koma9,Koma20}.
For example, energy exchange cosmology assumes the transfer of energy between two fluids \cite{Barrow22}, such as the interaction between dark matter and dark energy \cite{Wang0102}.
In this case, the two non-zero right-hand sides are totally cancelled because the total energy of the two fluids is conserved \cite{Koma20}.
In the bulk viscous and CCDM models, an effective formulation can be obtained from an effective description for pressure, using a single fluid, as examined in the next paragraph.
(When $f_{\Lambda} (t) =  h_{\textrm{B}} (t) =0$, the general cosmological equations reduce to those for standard cosmology and, therefore, the continuity equation is given by $\dot{\rho} + 3  H [ \rho +  (p/c^2)] =0$.    
The same continuity equation can be obtained when both $f_{\Lambda} (t) = \Lambda / 3$ and $ h_{\textrm{B}} (t) =0$ are considered, as for $\Lambda$CDM models.)

We now consider an effective formulation, using an effective pressure, $p_{e}= p + p^{\prime}$, which is given by \cite{Koma9}
\begin{align}
p_{e}(t)  &= p(t) + p^{\prime}(t) = p(t) - \frac{c^{2}}{4 \pi G}    h_{\textrm{B}}(t)     ,
\label{eq:p_e}
\end{align}
where $p^{\prime} (t)$ has been replaced by $ -  c^{2} h_{\textrm{B}}(t)/(4 \pi G)$.
Applying the effective pressure $p_{e}(t)$ to Eqs.\ (\ref{eq:General_FRW02}), (\ref{eq:drho_General}), and (\ref{eq:dotH}) yields \cite{Koma9}
\begin{align}
  \frac{ \ddot{a}(t) }{ a(t) }    &= -  \frac{ 4\pi G }{ 3 }  \left ( \rho (t) +  \frac{3 p_{e}(t)}{c^2} \right )                   +   f_{\Lambda}(t)  , 
\label{eq:General_FRW02_pe}
\end{align}
\begin{equation}
       \dot{\rho} + 3  H \left ( \rho (t) +  \frac{p_{e}(t)}{c^2} \right )       =    -  \frac{3}{8 \pi G }   \dot{f}_{\Lambda}(t)   , 
\label{eq:drho_General_pe}
\end{equation}
\begin{equation}
    \dot{H} = - 4\pi G  \left ( \rho (t) +  \frac{p_{e}(t)}{c^2} \right )     ,
\label{eq:dotH_pe}
\end{equation}
where $p_{e}(t)$ includes the $h_{\textrm{B}}(t)$ term.
The effective pressure $p_{e}(t)$ can be related to irreversible entropies because the $h_{\textrm{B}}(t)$ term is considered to be related to an irreversible entropy arising from dissipative processes, such as the bulk viscosity \cite{Koma7}.
The effective formulation is used in Sec.\ \ref{The first law}. 
The right-hand side of Eq.\ (\ref{eq:drho_General_pe}) is non-zero except when $\dot{f}_{\Lambda}(t)=0$.
This non-zero term affects the modified thermodynamics relation and is examined in Sec.\ \ref{The first law}.
The Friedmann equation for the effective formulation is given by Eq.\ (\ref{eq:General_FRW01}), because the Friedmann equation does not include $h_{\textrm{B}}(t)$.

It should be noted that coupling Eq.\ (\ref{eq:General_FRW01}) with Eq.\ (\ref{eq:General_FRW02}) yields the cosmological equation \cite{Koma14,Koma15,Koma16}, given by 
\begin{equation}
    \dot{H} = - \frac{3}{2} (1+w)  H^{2}  +  \frac{3}{2}   (1+w)  f_{\Lambda}(t)     + h_{\textrm{B}}(t)   ,
\label{eq:Back2}
\end{equation}
where $w$ represents the equation of the state parameter for a generic component of matter, which is given as $w = p/(\rho  c^2)$.
For a matter-dominated universe and a radiation-dominated universe, the values of $w$ are $0$ and $1/3$, respectively.
Instead of $p_{e}$, $p$ is used for $w$ because Eq.\ (\ref{eq:Back2}) includes $h_{\textrm{B}}(t)$.
Equation\ (\ref{eq:Back2}) can describe background evolutions of the universe in various cosmological models \cite{Koma14,Koma15,Koma16}.
Accordingly, Eq.\ (\ref{eq:Back2}) is used for the discussion of cosmological models, such as a constant $T_{\rm{KH}}$ model \cite{Koma19}, as examined later.

\subsection{Entropy $S_{H}$ and temperature $T_{H}$ on the horizon} 
\label{Entropy and temperature}

The horizon thermodynamics is closely related to the holographic principle \cite{Hooft-Bousso}, which assumes that the horizon of the universe has an associated entropy and an approximate temperature \cite{EassonCai}.
The entropy $S_{H}$ and the temperature $T_{H}$ on the Hubble horizon are introduced according to previous works \cite{Koma11,Koma12,Koma17,Koma18,Koma19,Koma20}.

We select the Bekenstein--Hawking entropy $S_{\rm{BH}}$ as the associated entropy because it is the most standard.
In general, the cosmological horizon is examined by replacing the event horizon of a black hole by the cosmological horizon \cite{Koma17,Koma18}. 
This replacement method has been widely accepted \cite{Jacob1995,Padma2010,Verlinde1,HDE,Padma2012AB,Cai2012,Moradpour,Hashemi,Wang,Padmanabhan2004,ShuGong2011,Koma14,Koma15,Koma16,Koma17,Koma19,Koma20} and we use it here.

Based on the form of the Bekenstein--Hawking entropy, the entropy $S_{\rm{BH}}$ is written as \cite{Hawking1Bekenstein1}  
\begin{equation}
S_{\rm{BH}}  = \frac{ k_{B} c^3 }{  \hbar G }  \frac{A_{H}}{4}   ,
\label{eq:SBH}
\end{equation}
where $k_{B}$ and $\hbar$ are the Boltzmann constant and the reduced Planck constant, respectively.
The reduced Planck constant is defined by $\hbar \equiv h/(2 \pi)$, where $h$ is the Planck constant \cite{Koma11,Koma12}.
$A_{H}$ is the surface area of the sphere with a Hubble horizon (radius) $r_{H}$ given by
\begin{equation}
     r_{H} = \frac{c}{H}   .
\label{eq:rH}
\end{equation}
Substituting $A_{H}=4 \pi r_{H}^2 $ into Eq.\ (\ref{eq:SBH}) and applying Eq.\ (\ref{eq:rH}) yields
\begin{equation}
S_{\rm{BH}}  = \frac{ k_{B} c^3 }{  \hbar G }   \frac{A_{H}}{4}       
                  =  \left ( \frac{ \pi k_{B} c^5 }{ \hbar G } \right )  \frac{1}{H^2}  
                  =    \frac{K}{H^2}    , 
\label{eq:SBH2}      
\end{equation}
where $K$ is a positive constant given by
\begin{equation}
  K =  \frac{  \pi  k_{B}  c^5 }{ \hbar G } . 
\label{eq:K-def}
\end{equation}
When a de Sitter universe is considered, $S_{\rm{BH}}$ is constant although the scale factor exponentially increases with time.
Differentiating Eq.\ (\ref{eq:SBH2}) with respect to $t$ yields the first derivative of $S_{\rm{BH}}$, given by \cite{Koma11,Koma12}
\begin{equation}
\dot{S}_{\rm{BH}}  
                          = \frac{d}{dt}   \left ( \frac{K}{H^{2}} \right )  =  \frac{-2K \dot{H} }{H^{3}}                  .
\label{eq:dSBH}      
\end{equation}
The second law of thermodynamics on the horizon, $\dot{S}_{\rm{BH}} \geq 0$, is satisfied in favored cosmological models, such as $\Lambda$CDM models \cite{Koma14}.
In the present study, the form of the Bekenstein--Hawking entropy, $S_{\rm{BH}}$, is typically used for the entropy $S_{H}$ on the cosmological horizon. 
(Various forms of black-hole entropy such as nonextensive entropy have been proposed \cite{Das2008,Radicella2010,LQG2004_123,Tsallis2012,Czinner1Czinner2,Barrow2020,Nojiri2022,Gohar2023}, as described in Refs.\ \cite{Koma18,Koma19,Koma20}.
For a general form of entropy related to the Friedmann equation, see, e.g., Ref.\ \cite{Nojiri2024}.)

Next, we introduce an approximate temperature $T_{H}$ on the Hubble horizon.
Before introducing a dynamical temperature, we review the Gibbons--Hawking temperature $T_{\rm{GH}}$, which is given by \cite{GibbonsHawking1977} 
\begin{equation}
T_{\rm{GH}}  = \frac{ \hbar H}{   2 \pi  k_{B}  }   .
\label{eq:T_GH}
\end{equation}
This equation indicates that $T_{\rm{GH}}$ is proportional to $H$ and is constant during the evolution of de Sitter universes \cite{Koma17,Koma19}.
In fact, $T_{\rm{GH}}$ is obtained from field theory in the de Sitter space \cite{GibbonsHawking1977}.
However, most universes are not pure de Sitter universes in that their horizons are dynamic.
That is, horizons of universes (including our Universe) are generally considered to be dynamic \cite{Koma19,Koma20}.
Therefore, we introduce a dynamical Kodama--Hayward temperature, based on a previous work \cite{Koma19}.

In fact, a similar dynamic horizon for black holes has been examined in the works of Hayward \cite{Dynamical-T-1998} and Hayward \textit{et al}. \cite{Dynamical-T-2008}, as described in Ref.\ \cite{Koma19}.
Hayward suggested a dynamical temperature on a black hole horizon and clarified the relationship between the surface gravity and the temperature on a dynamic apparent horizon for the Kodama observer \cite{Dynamical-T-1998}.
The Kodama--Hayward temperature on the cosmological horizon of an FLRW universe has been proposed \cite{Dynamical-T-2007,Dynamical-T-20092014}, based on the works of Hayward \textit{et al}. \cite{Dynamical-T-1998,Dynamical-T-2008}.

The Kodama--Hayward temperature $T_{\rm{KH}}$ for a flat FLRW universe can be written as \cite{Tu2018,Tu2019}
\begin{equation}
 T_{\rm{KH}} = \frac{ \hbar H}{   2 \pi  k_{B}  }  \left ( 1 + \frac{ \dot{H} }{ 2 H^{2} }\right )  .
\label{eq:T_KH}
\end{equation}
Here $H>0$ and $1 + \frac{ \dot{H} }{ 2 H^{2} } \ge 0$ are assumed for a non-negative temperature in an expanding universe \cite{Koma19,Koma20}.
When de Sitter universes are considered, $T_{\rm{KH}}$ reduces to $T_{\rm{GH}}$ because $\dot{H}=0$ and, therefore, $T_{\rm{KH}}$ is interpreted as an extended version of $T_{\rm{GH}}$ \cite{Koma19,Koma20}.
In the present paper, the Kodama--Hayward temperature $T_{\rm{KH}}$ is typically used for the temperature $T_{H}$ on the horizon.
Cosmological models used later satisfy a non-negative temperature.

As examined in Refs.\ \cite{Koma19,Koma20}, $T_{\rm{KH}}$ is constant when the following equation is satisfied:
\begin{align}
    \dot{H} &= - 2 H^{2}  +  2 \psi H_{0}  H  ,
\label{Cosmo_T_H_mod_cst}
\end{align}
where $\psi$ represents a dimensionless constant. 
Substituting Eq.\ (\ref{Cosmo_T_H_mod_cst}) into Eq.\ (\ref{eq:T_KH}) gives a constant temperature, $T_{\rm{KH}} = \hbar \psi H_{0} /(2 \pi k_{B})$ \cite{Koma19,Koma20}.
Accordingly, $T_{\rm{KH}}$ is non-negative when $\psi \ge 0$ is considered in an expanding universe.
A universe at constant $T_{\rm{KH}}$ has been studied, using a cosmological model which includes a power-law term proportional to $H^{\alpha}$ (where $\alpha$ is a free variable) and the equation of state parameter $w$ \cite{Koma19,Koma20}.
For example, substituting $w=1/3$, $f_{\Lambda} (t) = 0$, and $h_{\textrm{B}} (t) = 2 \psi H_{0}  H$ into Eq.\ (\ref{eq:Back2}) yields the cosmological equation equivalent to Eq.\ (\ref{Cosmo_T_H_mod_cst}).
(The $H$ term for $h_{\textrm{B}} (t)$ is obtained from $H^{\alpha}$ when $\alpha =1$.)
In this universe, $T_{\rm{KH}}$ is constant though the Hubble parameter varies with time.
Understanding the universe at constant $T_{\rm{KH}}$ should contribute to the study of horizon thermodynamics because systems at constant temperature play important roles in thermodynamics.
In Sec.\ \ref{Equipartition law}, we consider a constant $T_{\rm{KH}}$ universe, to simplify the modified thermodynamic relation.
In Sec.\ \ref{A constant TKH model}, the evolution of the constant $T_{\rm{KH}}$ universe is observed, using a constant $T_{\rm{KH}}$ model.
To formulate the constant $T_{\rm{KH}}$ model, a power-law model with $H^{\alpha}$ terms is discussed in Sec.\ \ref{A constant TKH model}.

We calculate $T_{\rm{GH}}  \dot{S}_{\rm{BH}}$ and $T_{\rm{KH}}  \dot{S}_{\rm{BH}}$, used to discuss the modified thermodynamic relation.
Substituting Eqs.\ (\ref{eq:T_GH}) and (\ref{eq:dSBH}) into $T_{\rm{GH}}  \dot{S}_{\rm{BH}}$ yields 
\begin{align}
T_{\rm{GH}}  \dot{S}_{\rm{BH}}  
                                             &=  \frac{ \hbar H}{   2 \pi  k_{B}  }   \left ( \frac{-2 \left ( \frac{  \pi  k_{B}  c^5 }{ \hbar G } \right ) \dot{H} }{H^{3}}   \right )            \notag \\
                                             &=  \left ( \frac{c^5 }{G} \right ) \left ( \frac{ - \dot{H} }{H^{2}} \right )     ,
\label{eq:TGH-dSBH}      
\end{align}
where $ K = \pi k_{B} c^5 /(\hbar G)$ given by Eq.\ (\ref{eq:K-def}) has been used.
Similarly, from Eqs.\ (\ref{eq:T_KH}) and (\ref{eq:dSBH}), $T_{\rm{KH}}  \dot{S}_{\rm{BH}}$ is given by 
\begin{align}
T_{\rm{KH}}  \dot{S}_{\rm{BH}}  
                                            &=  \left ( \frac{c^5 }{G} \right ) \left ( \frac{ - \dot{H} }{H^{2}} \right )  \left ( 1 + \frac{ \dot{H} }{ 2 H^{2} } \right )     .
\label{eq:TKH-dSBH}       
\end{align}
The obtained $T_{\rm{GH}}  \dot{S}_{\rm{BH}}$ and $T_{\rm{KH}}  \dot{S}_{\rm{BH}}$ are used later.

\section{First law of thermodynamics and the modified thermodynamic relation} 
\label{The first law}

In this section, we derive the modified thermodynamic relation by applying the general formulation for cosmological equations to the first law of thermodynamics.
For this, we review the first law according to Refs.\ \cite{Cai2007,Cai2007B,Cai2008,Sanchez2023,Nojiri2024,Odintsov2023ab,Odintsov2024,Odintsov2024B}.
In the present paper, a flat FLRW universe is considered and, therefore, the Hubble horizon is equivalent to an apparent horizon.

The first law of thermodynamics is written as \cite{Cai2007,Cai2007B,Cai2008,Sanchez2023,Nojiri2024,Odintsov2023ab,Odintsov2024,Odintsov2024B}
\begin{align}
-dE_{\rm{bulk}}    + W dV  &  = T_{H} dS_{H}  ,
\label{eq:1stLaw}      
\end{align}
where $E_{\rm{bulk}}$ is the total internal energy of the matter fields inside the horizon, given by
\begin{align}
E_{\rm{bulk}}  &= \rho c^{2} V .
\label{Ebulk}
\end{align}
$W$ represents the work density done by the matter fields \cite{Nojiri2024}, which is written as
\begin{align}
W  &= \frac{\rho c^{2} - p }{2}   ,
\label{eq:W1}      
\end{align}
and $V$ is the Hubble volume, written as
\begin{equation}
V = \frac{4 \pi}{3} r_{H}^{3} =  \frac{4 \pi}{3} \left ( \frac{c}{H} \right )^{3}   ,
\label{eq:V}
\end{equation}
where $r_{H} = c/H$ given by Eq.\ (\ref{eq:rH}) is used.
Equation\ (\ref{eq:1stLaw}) indicates that the entropy on the horizon is generated, based on both the decreasing total internal energy of the bulk ($-dE_{\rm{bulk}}$) and the work done by the matter fields ($WdV$) \cite{Nojiri2024}.
The right-hand side of Eq.\ (\ref{eq:1stLaw}) should correspond to thermodynamic quantities on the horizon and may be modified as $T_{H} dS_{e}$, using an effective entropy $S_{e}$ \cite{Cai2007B}, as discussed later.
In contrast, the left-hand side may be related to be the energy flux of matter fields from inside to outside the horizon for an infinitesimal time, as described in Ref.\ \cite{Nojiri2024}.
The work density $W=(\rho c^{2} - p)/2$ given by Eq.\ (\ref{eq:W1}) is based on the fact that the work done is obtained by the trace of the energy-momentum tensor of the matter fields along the direction perpendicular to the (apparent) horizon \cite{Odintsov2024B}.
We note that an alternative work density is discussed in Ref.\ \cite{Nojiri2024}.

In addition, Eq.\ (\ref{eq:1stLaw}) can be written as 
\begin{align}
- \frac{dE_{\rm{bulk}}}{dt}    + W \frac{dV}{dt}  &= T_{H} \frac{dS_{H}}{dt}  ,
\label{eq:1stLaw_dt0}      
\end{align}
or equivalently, 
\begin{align}
-\dot{E}_{\rm{bulk}}    + W \dot{V}  &=  T_{H} \dot{S}_{H}    .
\label{eq:1stLaw_dt1}      
\end{align}
The left-hand side describes thermodynamic quantities in the bulk, whereas the right-hand side describes thermodynamic quantities on the horizon.
Originally, $S_{H} = S_{\rm{BH}}$ and $T_{H} = T_{\rm{KH}}$ were considered \cite{Cai2007} and the first law is satisfied in standard cosmology.

In this study, we typically use $S_{H} = S_{\rm{BH}}$ and $T_{H} = T_{\rm{KH}}$.
However, $S_{H}$ and $T_{H}$ are retained so that we can consider other entropies and temperatures.
For example, when nonextensive entropies \cite{Das2008,Radicella2010,LQG2004_123,Tsallis2012,Czinner1Czinner2,Barrow2020,Nojiri2022,Gohar2023} are used for $S_{H}$, the first law of thermodynamics should lead to the modified Friedmann and acceleration equations, as examined in previous works; see, e.g., Ref.\ \cite{Odintsov2024}.
Also, a general form of entropy has been discussed based on the first law \cite{Nojiri2024}.

We examine Eq.\ (\ref{eq:1stLaw_dt1}) and calculate the left-hand side of this equation.
Substituting Eqs.\ (\ref{Ebulk}) and  (\ref{eq:W1}) into $-\dot{E}_{\rm{bulk}}    + W \dot{V}$ yields \cite{Nojiri2024}
\begin{align}
-\dot{E}_{\rm{bulk}}    + W \dot{V}  &= - \frac{d (\rho c^{2} V)}{dt}                    + \left ( \frac{\rho c^{2} - p }{2} \right )  \dot{V}    \notag \\
                                                             &= -\dot{\rho} c^{2} V  - \rho c^{2} \dot{V}  + \left ( \frac{\rho c^{2} - p }{2} \right )  \dot{V}    \notag \\
                                                             &= -\dot{\rho} c^{2} V  - \left ( \frac{\rho c^{2} + p }{2} \right )  \dot{V}    .
\label{eq:Left1stLaw}      
\end{align}
The above equation indicates that  $-\dot{E}_{\rm{bulk}}   + W \dot{V}$ is divided into a $\dot{\rho}$ term and a $\dot{V}$ term, namely $-\dot{\rho} c^{2} V$ and $ - [(\rho c^{2} + p)/2] \dot{V}$. 
Hereafter we call the two terms the `$\dot{\rho}$ and $\dot{V}$ terms', even when $p$ is replaced by $p_{e}$.
This result implies that $T_{\rm{KH}}  \dot{S}_{\rm{BH}}$ can also be divided into two parts, corresponding to the $\dot{\rho}$ and $\dot{V}$ terms.
This consistency is examined in Sec.\ \ref{drho and dV terms}.
Note that the $\dot{V}$ term is the sum of $- \rho c^{2} \dot{V}$ and $W \dot{V}$.

We now apply general cosmological equations to the first law of thermodynamics.
Equation\ (\ref{eq:Left1stLaw}) corresponds to the left-hand side of Eq.\ (\ref{eq:1stLaw_dt1}).
$\dot{\rho}$ and $\rho c^{2} + p$ in Eq.\ (\ref{eq:Left1stLaw}) are calculated using Eqs.\ (\ref{eq:drho_General}) and (\ref{eq:dotH}).
Substituting Eqs.\ (\ref{eq:drho_General}) and (\ref{eq:dotH}) into Eq.\ (\ref{eq:Left1stLaw}), applying $V= (4 \pi/3)(c/H)^{3}$ given by Eq.\ (\ref{eq:V}) and $\dot{V}= -4 \pi c^{3} H^{-4} \dot{H}$, and performing several operations yields
\begin{widetext}
\begin{align}
     -\dot{E}_{\rm{bulk}}    + W \dot{V}   =  -\dot{\rho} c^{2} V  - \left ( \frac{\rho c^{2} + p }{2} \right )  \dot{V}   
                                                         &=   \left ( \frac{c^5 }{G} \right ) \left ( \frac{ - \dot{H} }{H^{2}} \right )  \left ( 1 + \frac{ \dot{H} }{ 2 H^{2} } \right )            
                                                              + \frac{1}{2} \left ( \frac{c^5 }{G} \right ) \left ( \frac{ \dot{f}_{\Lambda}(t)  }{H^{3}}  + \frac{ \dot{H} h_{\textrm{B}}(t) }{ H^{4} } \right )     \notag \\                  
                                                         &=   T_{\rm{KH}}  \dot{S}_{\rm{BH}}                 
                                                             + \frac{1}{2} \left ( \frac{c^5 }{G} \right ) \left ( \frac{ \dot{f}_{\Lambda}(t)  }{H^{3}}  + \frac{ \dot{H} h_{\textrm{B}}(t) }{ H^{4} } \right )      , 
\label{eq:Left1stLaw_2}      
\end{align}
\end{widetext}
where $T_{\rm{KH}}  \dot{S}_{\rm{BH}}$ given by Eq.\ (\ref{eq:TKH-dSBH}) is also used.
Equation\ (\ref{eq:Left1stLaw_2}) corresponds to the left-hand side of Eq.\ (\ref{eq:1stLaw_dt1}).
Therefore, from Eq.\ (\ref{eq:Left1stLaw_2}), the first law based on Eq.\ (\ref{eq:1stLaw_dt1}) may be written as
\begin{align}
     -\dot{E}_{\rm{bulk}}    + W \dot{V}                
                                                         &=   T_{\rm{KH}}  \dot{S}_{\rm{BH}}                 
                                                             + \frac{1}{2} \left ( \frac{c^5 }{G} \right ) \left ( \frac{ \dot{f}_{\Lambda}(t)  }{H^{3}}  + \frac{ \dot{H} h_{\textrm{B}}(t) }{ H^{4} } \right )      \notag \\
                                                        &=T_{H} \dot{S}_{e}                                                                                                                                                                                     \notag \\
                                                        &=T_{H} \dot{S}_{H}  + T_{H} (\dot{S}_{\dot{f}_{\Lambda}}  + \dot{S}_{h_{\textrm{B}}} )  ,
\label{eq:1stLaw_general}      
\end{align}
where ${S}_{e}$ represents an effective entropy \cite{Cai2007B}.
In this study, ${S}_{e}$ is assumed to be given by 
\begin{align}
   S_{e} = S_{H} +  S_{\dot{f}_{\Lambda}} + S_{h_{\textrm{B}}}   ,
\label{eq:Seff}      
\end{align}
where $S_{\dot{f}_{\Lambda}}$ and $S_{h_{\textrm{B}}}$ represent reversible and irreversible entropies, which are related to the additional $\dot{f}_{\Lambda} (t)$ and $h_{\textrm{B}}(t)$ terms in the first line of Eq.\ (\ref{eq:1stLaw_general}), respectively.
Equation\ (\ref{eq:1stLaw_general}) is considered to be a generalized first law of thermodynamics.
We call this a `modified thermodynamic relation', because Eq.\ (\ref{eq:1stLaw_general}) has not yet been established.
We can confirm that when $\dot{f}_{\Lambda} (t) = h_{\textrm{B}}(t) =0$, Eq.\ (\ref{eq:1stLaw_general}) reduces to $-\dot{E}_{\rm{bulk}}    + W \dot{V}  =  T_{\rm{KH}}  \dot{S}_{\rm{BH}} =T_{H} \dot{S}_{H}$, where $\dot{S}_{\dot{f}_{\Lambda}} = \dot{S}_{h_{\textrm{B}}} =0$ is also used.

The additional $\dot{f}_{\Lambda} (t)$ and $h_{\textrm{B}}(t)$ terms in Eq.\ (\ref{eq:1stLaw_general}) are based on two extra driving terms included in general cosmological equations, and these two terms are assumed to be related to entropies.
In this paper, the $\dot{f}_{\Lambda} (t)$ term is considered to be related to the reversible entropy $S_{\dot{f}_{\Lambda}}$, such as that related to the reversible exchange of energy. 
The $h_{\textrm{B}}(t)$ term is considered to be related to the irreversible entropy $S_{h_{\textrm{B}}}$ arising from dissipative processes, such as the bulk viscosity and the creation of CDM \cite{Koma7}.
Other entropies, such as nonequilibrium entropies and generalized entropies, are not considered here.
The relationship between similar additional terms and nonequilibrium entropies has been discussed in, e.g., Refs.\ \cite{Dynamical-T-2007,Cai2007B,Santos2022}.
A general form of entropy that connects the Friedmann equations for any gravity theory with horizon thermodynamics has been examined in Ref.\ \cite{Nojiri2024}.
We expect that modified FLRW equations can be derived from Eq.\ (\ref{eq:1stLaw_general}), using various forms of entropy $S_{H}$ on the horizon \cite{Das2008,Radicella2010,LQG2004_123,Tsallis2012,Czinner1Czinner2,Barrow2020,Nojiri2022,Gohar2023}.
The modified thermodynamic relation given by Eq.\ (\ref{eq:1stLaw_general}) should provide new insights into thermodynamics and cosmological equations.

For an effective formulation, we apply an effective pressure $p_{e}$ given by Eq.\ (\ref{eq:p_e}) to the modified thermodynamic relation.
Using $p_{e}$, Eq.\ (\ref{eq:1stLaw_general}) can be written as 
\begin{align}
     -\dot{E}_{\rm{bulk}}    + W_{e} \dot{V}   &=  T_{\rm{KH}}  \dot{S}_{\rm{BH}}  + \frac{1}{2} \left ( \frac{c^5 }{G} \right ) \left ( \frac{ \dot{f}_{\Lambda}(t)  }{H^{3}} \right )  \notag \\
                                                              &=  T_{H} \dot{S}_{H}   + T_{H} \dot{S}_{\dot{f}_{\Lambda}}   ,
\label{eq:1stLaw_general_pe}      
\end{align}
where $W_{e}$ is the effective work density, given by
\begin{align}
  W_{e}   &= \frac{\rho c^{2} - p_{e} }{2}   . 
\label{eq:W1_pe}      
\end{align}
Equation\ (\ref{eq:1stLaw_general_pe}) is also a `modified thermodynamic relation'.
$p_{e}$ includes the effect of $S_{h_{\textrm{B}}}$.
The additional $\dot{S}_{\dot{f}_{\Lambda}}$ and $\dot{f}_{\Lambda}(t)$ terms should lead to modified cosmological equations and may be related to a corrected term for a generalized entropy.
These tasks are left for future research.
Hereafter, we consider a constant $f_{\Lambda}(t)$ and neglect the additional terms.
Substituting $\dot{f}_{\Lambda} (t) =0$ into Eq.\ (\ref{eq:1stLaw_general_pe}) yields 
\begin{align}
-\dot{E}_{\rm{bulk}}    + W_{e} \dot{V}  =  T_{\rm{KH}}  \dot{S}_{\rm{BH}} = T_{H} \dot{S}_{H}  , 
\label{eq:modified1stLaw}    
\end{align}
where $\dot{S}_{\dot{f}_{\Lambda}} =0$ has been also used. 
This simplified relation implicitly includes $p_{e}$ and a constant $f_{\Lambda}(t)$, corresponding to a cosmological constant.
Equation\ (\ref{eq:modified1stLaw}) has not yet been established, but it is consistent with the formulation of the first law given by Eq.\ (\ref{eq:1stLaw_dt1}).
When $p_{e}=p$, this relation is equivalent to the first law itself, because $W_{e}=W$.
In fact, Eq.\ (\ref{eq:modified1stLaw}) is expected to provide significant information even though the additional terms are neglected.
Therefore, we examine Eq.\ (\ref{eq:modified1stLaw}) in the next section.

\section{Modified thermodynamic relation for constant $f_{\Lambda}(t)$} 
\label{A modified thermodynamic relation}

In this section, we examine the modified thermodynamic relation for constant $f_{\Lambda}(t)$.
That is, we consider a constant $f_{\Lambda}(t)$ and an effective pressure $p_{e}$.
From Eq.\ (\ref{eq:modified1stLaw}), the modified thermodynamic relation is written as
\begin{align}
     -\dot{E}_{\rm{bulk}}    + W_{e} \dot{V}   &=  T_{\rm{KH}}  \dot{S}_{\rm{BH}}  =  T_{H} \dot{S}_{H}  ,   
\label{eq:1stLaw_general_pe_df=0}      
\end{align}
or equivalently,
\begin{align}
     -d{E}_{\rm{bulk}}    + W_{e} dV   &=  T_{\rm{KH}}  d{S}_{\rm{BH}}  =  T_{H} d{S}_{H}                      ,   
\label{eq:1stLaw_general_pe_df=0_dE}      
\end{align}
where $W_{e}$ is the effective work density $(\rho c^{2} - p_{e})/2$, given by Eq.\ (\ref{eq:W1_pe}).
The above two equations are the modified thermodynamic relation examined in this section. 
The left-hand sides of the two equations describe thermodynamic quantities in the bulk, while the middles and the right-hand sides describe thermodynamic quantities on the horizon.
In $S_{H}$ and $T_{H}$, the subscript $H$ indicates thermodynamic quantities on the horizon, as examined below.

The Helmholtz free energy $F$ is defined as $F = E - T S$ \cite{Callen} and, therefore, $dF$ is given by $dF = dE - TdS  - SdT$. 
Accordingly, $TdS$ and $T \dot{S}$ are written as
\begin{equation}
   TdS = dE - dF  - SdT,    \quad  T \dot{S} = \dot{E} - \dot{F}  - S\dot{T}  .
\label{TdS_thermo}
\end{equation}
Applying Eq.\ (\ref{TdS_thermo}) to the right-hand side of Eqs.\ (\ref{eq:1stLaw_general_pe_df=0}) and (\ref{eq:1stLaw_general_pe_df=0_dE}) yields the modified thermodynamic relation: 
\begin{align}
     -\dot{E}_{\rm{bulk}}    + W_{e} \dot{V}   =  T_{\rm{KH}}  \dot{S}_{\rm{BH}}  &=  T_{H} \dot{S}_{H}                                  \notag \\   
                                                                                                               &=   \dot{E}_{H} - \dot{F}_{H}   - S_{H} \dot{T}_{H}       ,
\label{eq:1stLaw_general_pe_df=0_dEdF}      
\end{align}
\begin{align}
     -d{E}_{\rm{bulk}}    + W_{e} dV    =  T_{\rm{KH}}  d{S}_{\rm{BH}}  &=  T_{H} d{S}_{H}                                                       \notag \\   
                                                                                               &=   dE_{H} - dF_{H}   - S_{H} dT_{H}       ,           
\label{eq:1stLaw_general_pe_df=0_dEdF_2}      
\end{align}
where $E_{H}$ and $F_{H}$ represent an energy and the Helmholtz free energy on the horizon, respectively.

As examined in Sec.\ \ref{The first law}, Eq.\ (\ref{eq:Left1stLaw}) implies that the left-hand side of the modified thermodynamic relation can be divided into `$\dot{\rho}$ and $\dot{V}$ terms'.
Similarly, it is expected that $T_{H} \dot{S}_{H}$ ($=T_{\rm{KH}}  \dot{S}_{\rm{BH}}$) can be divided into two parts, corresponding to the $\dot{\rho}$ and $\dot{V}$ terms.
In addition, thermodynamic quantities on the horizon, namely the right-hand side of the modified thermodynamic relation, can be further simplified by applying the equipartition law of energy on the horizon.
Accordingly, in Sec.\ \ref{drho and dV terms}, the $\dot{\rho}$ and $\dot{V}$ terms are examined.
In Sec.\ \ref{Equipartition law}, the equipartition law is applied to the modified thermodynamic relation.

\subsection{$\dot{\rho}$ and $\dot{V}$ terms: thermodynamic quantities in the bulk for the modified thermodynamic relation} 
\label{drho and dV terms}

In this subsection, we examine the $\dot{\rho}$ and $\dot{V}$ terms in the modified thermodynamic relation for constant $f_{\Lambda}(t)$.
The $\dot{\rho}$ and $\dot{V}$ terms correspond to the left-hand side of Eq.\ (\ref{eq:1stLaw_general_pe_df=0}), as confirmed later.
The sum of the two terms is equivalent to $T_{\rm{KH}}  \dot{S}_{\rm{BH}}$, from Eq.\ (\ref{eq:1stLaw_general_pe_df=0}).
Accordingly, we first discuss $T_{\rm{KH}}  \dot{S}_{\rm{BH}}$ and then examine the $\dot{\rho}$ and $\dot{V}$ terms.

Using $T_{\rm{KH}}$ and $T_{\rm{GH}}$, $T_{\rm{KH}}  \dot{S}_{\rm{BH}}$ can be written as
\begin{align}
      T_{\rm{KH}}  \dot{S}_{\rm{BH}}  =  T_{\rm{GH}}  \dot{S}_{\rm{BH}}  + \left (\frac{T_{\rm{KH}}}{T_{\rm{GH}}} - 1   \right ) T_{\rm{GH}}  \dot{S}_{\rm{BH}}   ,
\label{eq_TKHdSBH_extend}      
\end{align}
where $T_{\rm{GH}}$ is the Gibbons--Hawking temperature given by Eq.\ (\ref{eq:T_GH}).
The first term on the right-hand side of Eq.\ (\ref{eq_TKHdSBH_extend}) is $T_{\rm{GH}}  \dot{S}_{\rm{BH}}$.
The second term is equivalent to the product of $T_{\rm{GH}}  \dot{S}_{\rm{BH}}$ and a normalized deviation of $T_{\rm{KH}}$ from $T_{\rm{GH}}$.
In de Sitter universes, this deviation reduces to $0$ because $T_{\rm{KH}} = T_{\rm{GH}}$.
From Eq.\ (\ref{eq:TGH-dSBH}),  $T_{\rm{GH}}  \dot{S}_{\rm{BH}}$ is given by 
\begin{align}
T_{\rm{GH}}  \dot{S}_{\rm{BH}}     &=  \left ( \frac{c^5 }{G} \right ) \left ( \frac{ - \dot{H} }{H^{2}} \right )     .
\label{eq:TGH-dSBH_2}      
\end{align}
Using Eqs.\ (\ref{eq:T_GH}), (\ref{eq:T_KH}), and (\ref{eq:TGH-dSBH_2}), we have the second term on the right-hand side of Eq.\ (\ref{eq_TKHdSBH_extend}), written as
\begin{align}
\left (\frac{T_{\rm{KH}}}{T_{\rm{GH}}} - 1   \right ) T_{\rm{GH}}  \dot{S}_{\rm{BH}} &= \left ( \frac{  \dot{H} }{ 2H^{2}} \right )  \left ( \frac{c^5 }{G} \right ) \left ( \frac{ - \dot{H} }{H^{2}} \right )  .
\label{eq:2nd-term}      
\end{align}
Coupling Eq.\ (\ref{eq:TGH-dSBH_2}) with Eq.\ (\ref{eq:2nd-term}) yields
\begin{align}
\left (\frac{T_{\rm{KH}}}{T_{\rm{GH}}} - 1   \right ) T_{\rm{GH}}  \dot{S}_{\rm{BH}} &=  - \frac{1}{2 (c^5/G)}  \left ( T_{\rm{GH}}  \dot{S}_{\rm{BH}}  \right )^{2}   .
\label{eq:1st-2nd-terms}      
\end{align}
In addition, normalizing Eqs.\ (\ref{eq:TGH-dSBH_2}) and (\ref{eq:2nd-term}) by $T_{\rm{GH},0} (S_{\rm{BH},0} H_{0})$ and coupling the two resulting equations yields
\begin{align}
   \frac{ \left (\frac{T_{\rm{KH}}}{T_{\rm{GH}}} - 1   \right ) T_{\rm{GH}}  \dot{S}_{\rm{BH}} }{ T_{\rm{GH},0}  (S_{\rm{BH},0} H_{0}) }   &=   - \frac{1}{4}   \left [   \frac{ T_{\rm{GH}}  \dot{S}_{\rm{BH}} }{ T_{\rm{GH},0}  (S_{\rm{BH},0} H_{0}) } \right ]^{2}    , 
\label{eq:1st-2nd-terms_Norm}      
\end{align}
where $T_{\rm{GH},0} =  \hbar H_{0} / (2 \pi  k_{B})$, $S_{\rm{BH},0} =  K/H_{0}^{2}$, and $K =  \pi  k_{B} c^5/(\hbar G)$ have been used.
$H_{0}$ represents the Hubble parameter at the present time.
Equations\ (\ref{eq:1st-2nd-terms}) and (\ref{eq:1st-2nd-terms_Norm}) indicate that the magnitude of the second term is proportional to the square of the first term.

We now calculate $-\dot{E}_{\rm{bulk}}    + W_{e} \dot{V}$, namely the left-hand side of Eq.\ (\ref{eq:1stLaw_general_pe_df=0}).
Using Eq.\ (\ref{eq:Left1stLaw}) and replacing $p$ and $W$ by $p_{e}$ and $W_{e}$, respectively, yields
\begin{align}
-\dot{E}_{\rm{bulk}}    + W_{e} \dot{V}        &= -\dot{\rho} c^{2} V  - \left ( \frac{\rho c^{2} + p_{e} }{2} \right )  \dot{V}    ,
\label{eq:Left1stLaw_pe_df=0}      
\end{align}
where $p_{e}$ is an effective pressure given by Eq.\ (\ref{eq:p_e}) and $W_{e}$ is an effective work density $(\rho c^{2} - p_{e})/2$, given by Eq.\ (\ref{eq:W1_pe}).
In this way, we confirm that $-\dot{E}_{\rm{bulk}}    + W_{e} \dot{V}$ can be divided into $-\dot{\rho} c^{2} V$ and $ - [(\rho c^{2} + p_{e})/2] \dot{V}$, namely the `$\dot{\rho}$ and $\dot{V}$ terms'.
The $\dot{V}$ term is the sum of $- \rho c^{2} \dot{V}$ and $W_{e} \dot{V}$, where $- \rho c^{2} \dot{V}$ is implicitly included in $-\dot{E}_{\rm{bulk}}$.
As examined previously, $\dot{\rho}$ can be replaced by applying the continuity equation $\dot{\rho} + 3  H  [ \rho + (p_{e}/c^2)]  = 0$ given by Eq.\ (\ref{eq:drho_General_pe}) and $ \dot{H} = - 4\pi G [ \rho + (p_{e}/c^2)] $ given by Eq.\ (\ref{eq:dotH_pe}), where $\dot{f}_{\Lambda}(t) =0$ has been used.
From these equations, we obtain the first term on the right-hand side of Eq.\ (\ref{eq:Left1stLaw_pe_df=0}), namely $-\dot{\rho} c^{2} V$, which can be written as
\begin{align}
-\dot{\rho} c^{2} V    &= - \left ( - 3  H  \left ( \rho + \frac{p_{e}}{c^2} \right ) \right )  c^{2} V   \notag \\
                               &=   3 H  \left ( - \frac{\dot{H}}{4\pi G}  \right )  c^{2} \frac{4 \pi}{3} \left ( \frac{c}{H} \right )^{3}  
                                  = \left ( \frac{c^5 }{G} \right ) \left ( \frac{ - \dot{H} }{H^{2}} \right )  , 
\label{eq:drho_pe_df=0}      
\end{align}
where $V= (4 \pi/3)(c/H)^{3}$ given by Eq.\ (\ref{eq:V}) has been used.
Equation\ (\ref{eq:drho_pe_df=0}) is equivalent to Eq.\ (\ref{eq:TGH-dSBH_2}).
In addition, substituting both $ \dot{H} = - 4\pi G [ \rho + (p_{e}/c^2)] $ and  $\dot{V}= -4 \pi c^{3} H^{-4} \dot{H}$ into the second term on the right-hand side of Eq.\ (\ref{eq:Left1stLaw_pe_df=0}) yields
\begin{align}
- \left ( \frac{\rho c^{2} + p_{e} }{2} \right )  \dot{V}  &=    - \frac{1}{2} \left ( - \frac{ \dot{H} c^{2} } {4\pi G} \right )        \left (\frac{-4 \pi c^{3}  \dot{H} }{H^{4}}  \right   )     \notag \\
                                                                             &= \left ( \frac{  \dot{H} }{ 2H^{2}} \right )  \left ( \frac{c^5 }{G} \right ) \left ( \frac{ - \dot{H} }{H^{2}} \right )  .
\label{eq:dV_pe_df=0}      
\end{align}
Equation\ (\ref{eq:dV_pe_df=0}) is equivalent to Eq.\ (\ref{eq:2nd-term}).
From these two results, we have the following two equations:
\begin{align}
-\dot{\rho} c^{2} V  &= T_{\rm{GH}}  \dot{S}_{\rm{BH}} , 
\label{eq:drho=TGHdSBH_pe_df=0}      
\end{align}
\begin{align}
- \left ( \frac{\rho c^{2} + p_{e} }{2} \right )  \dot{V}  &= \left (\frac{T_{\rm{KH}}}{T_{\rm{GH}}} - 1   \right ) T_{\rm{GH}}  \dot{S}_{\rm{BH}}  .
\label{eq:dV=2nd-term_pe_df=0}      
\end{align}
Therefore, the modified thermodynamic relation given by Eq.\ (\ref{eq:1stLaw_general_pe_df=0}) can be summarized as
\begin{widetext}
\begin{equation}
  -\dot{E}_{\rm{bulk}}    + W_{e} \dot{V}        =  \underbrace{-\dot{\rho} c^{2} V}_{T_{\rm{GH}}  \dot{S}_{\rm{BH}}}  \underbrace{-\- \left ( \frac{\rho c^{2} + p_{e} }{2} \right )  \dot{V}}_{\left (\frac{T_{\rm{KH}}}{T_{\rm{GH}}} - 1   \right ) T_{\rm{GH}}  \dot{S}_{\rm{BH}} }  
                                                                       = T_{\rm{KH}}  \dot{S}_{\rm{BH}}  =  T_{H} \dot{S}_{H} ,         
\label{eq:Left1stLaw_pe_df=0_drhodV}      
\end{equation}
\end{widetext}
where Eq.\ (\ref{eq:Left1stLaw_pe_df=0}) is also used.
Equation\ (\ref{eq:Left1stLaw_pe_df=0_drhodV}) indicates that the $\dot{\rho}$ term is equivalent to $T_{\rm{GH}}  \dot{S}_{\rm{BH}}$, whereas the $\dot{V}$ term is equivalent to $[(T_{\rm{KH}}/T_{\rm{GH}}) -1] T_{\rm{GH}}  \dot{S}_{\rm{BH}}$.
In this way, the modified thermodynamic relation can be interpreted as consisting of contributions from the $\dot{\rho}$ and $\dot{V}$ terms.
We expect that Eq.\ (\ref{eq:Left1stLaw_pe_df=0_drhodV}) should provide a better understanding of the modified thermodynamic relation.
For example, based on Eq.\ (\ref{eq:Left1stLaw_pe_df=0_drhodV}), Eq.\ (\ref{eq:1st-2nd-terms}) can be interpreted as showing that the magnitude of the $\dot{V}$ term is proportional to the square of the $\dot{\rho}$ term.
This relationship is satisfied even when $W_{e}=W$, namely the first law of thermodynamics.
In addition, solving $ -\dot{\rho} c^{2} V = T_{\rm{GH}}  \dot{S}_{\rm{BH}}$ given by Eq.\ (\ref{eq:Left1stLaw_pe_df=0_drhodV}) with respect to $\dot{\rho}$, 
substituting both $V= (4 \pi/3)(c/H)^{3}$ and $T_{\rm{GH}}  \dot{S}_{\rm{BH}}   =   ( c^5/G ) ( - \dot{H} / H^{2} )$ into the resulting equation 
and integrating yields the Friedmann equation, written as  
\begin{align}
 H^2 = \frac{ 8\pi G }{ 3 } \rho+ C, 
\label{eq:Friedmann_drho} 
\end{align}
where $C$ represents integral constants.
That is, the $\dot{\rho}$ term in the thermodynamic relation corresponds to the Friedmann equation.
Similarly, solving $ - [(\rho c^{2} + p_{e})/2] \dot{V}  = [(T_{\rm{KH}}/T_{\rm{GH}}) -1] T_{\rm{GH}}  \dot{S}_{\rm{BH}}$ with respect to $[\rho + (p_{e}/c^2)]$ and substituting Eq.\ (\ref{eq:2nd-term}) and $\dot{V}= -4 \pi c^{3} H^{-4} \dot{H}$ into the resulting equation 
gives $ \dot{H} = - 4\pi G [ \rho + (p_{e}/c^2)] $, which is equivalent to Eq.\ (\ref{eq:dotH_pe}).
In addition, adding this equation to the derived Friedmann equation yields an acceleration equation, written as  
\begin{align}
 \frac{\ddot{a}}{a} = \dot{H} + H^{2} = -\frac{ 4\pi G }{ 3 }  \left ( \rho +  \frac{3 p_e}{c^2} \right )   +  C  .
\end{align}
The derived Friedmann and acceleration equations are not unexpected because, in the previous section, the modified thermodynamic relation was similarly derived from general cosmological equations and the first law of thermodynamics.
However, as examined above, Eq.\ (\ref{eq:Left1stLaw_pe_df=0_drhodV}) can clarify the roles of the $\dot{\rho}$ and $\dot{V}$ terms included in the modified thermodynamic relation.
In Sec.\ \ref{A constant TKH model}, we will study the evolution of the two terms, using cosmological models such as $\Lambda$CDM models.

\subsection{Applying the equipartition law of energy on the horizon to the modified thermodynamics relation} 
\label{Equipartition law}

In the previous subsection, we examined the $\dot{\rho}$ and $\dot{V}$ terms, corresponding to the left-hand side of the modified thermodynamic relation for constant $f_{\Lambda}(t)$.
We expect that the right-hand side of this relation, namely $T_{H} \dot{S}_{H}$, reduces to $\dot{F}_{H}$ by applying the equipartition law of energy on the horizon, as if the holographic-like connection $E_{\rm{bulk}}=F_{H}$ is extended.
For this, we first review the equipartition law, based on previous works \cite{Koma18,Koma20}.
Then, we apply the equipartition law to the right-hand side of the modified thermodynamics relation.
The equipartition law used here has not yet been established in a cosmological spacetime but is considered to be a viable scenario.

We have assumed that information for the bulk is stored on the horizon based on the holographic principle.  
We now assume the equipartition law of energy on the horizon \cite{Padma2010,ShuGong2011}.
Consequently, an energy on the Hubble horizon, $E_{H}$, can be written as
\begin{equation}
E_{H} =  N_{\rm{sur}}  \left ( \frac{1}{2} k_{B} T_{H}  \right )    , 
\label{E_equip}
\end{equation}
where $N_{\rm{sur}}$ is the number of degrees of freedom on a spherical surface of the Hubble radius $r_{H}$ and is written as \cite{Koma14}
\begin{equation}
  N_{\rm{sur}} = \frac{4 S_{H} }{k_{B}}       .
\label{N_sur}
\end{equation}
Substituting Eq.\ (\ref{N_sur}) into Eq.\ (\ref{E_equip}) yields  \cite{Padmanabhan2004,Padma2010}
\begin{equation}
E_{H} =   \frac{4 S_{H} }{k_{B}}     \left ( \frac{1}{2} k_{B} T_{H}  \right )    =  2 S_{H}  T_{H}  . 
\label{E_ST2_thermo_2}
\end{equation}
The above relation ${E}_{H}   =2 S_{H}  T_{H} $ was proposed by Padmanabhan \cite{Padmanabhan2004,Padma2010}.
(The same relation can be obtained from Euclidean action \cite{GibbonsHawking1977Action,York1986,York1988,Whiting1990,Wei2022} and a general `action--entropy relation' \cite{Broglie,ActionEntropy}, as discussed in Appendix\ \ref{A consistency of the equipartition law}.)
Originally, $S_{H} = S_{\rm{BH}}$ and $T_{H} = T_{\rm{GH}}$ were considered \cite{Padmanabhan2004,Padma2010}.
That is, the Gibbons--Hawking temperature $T_{\rm{GH}}$ was used for $T_{H}$.
In that case, the thermodynamic relation on the horizon was given by $dE_{H} = T_{\rm{GH}} dS_{\rm{BH}}$.
Accordingly, $\dot{E}_{H}$ was equivalent to $-\dot{\rho} c^{2} V$.
This equivalence can be easily confirmed from Eq.\ (\ref{eq:drho=TGHdSBH_pe_df=0}) and $dE_{H} = T_{\rm{GH}} dS_{\rm{BH}}$.

In the present paper, the Kodama--Hayward temperature $T_{\rm{KH}}$ is used for $T_{H}$, i.e., we set $T_{H} = T_{\rm{KH}}$.
The dynamical temperature should be appropriate discussing the equipartition law of energy on a dynamic horizon, especially when a universe at constant $T_{H}$ is considered.
In fact, a constant $T_{H}$ universe whose Hubble volume varies with time has been examined in Refs.\ \cite{Koma19,Koma20}, using the Kodama--Hayward temperature.
A constant $T_{H}$ universe is appropriate for studying the thermodynamics on dynamic horizons, because the dynamical temperature is constant as for de Sitter universes \cite{Koma19}.
In de Sitter universes, $T_{\rm{KH}}$ reduces to $T_{\rm{GH}}$, because the Hubble parameter is constant.
In the next section, we examine the evolution of thermodynamic quantities for a constant $T_{\rm{KH}}$ universe, using a constant $T_{\rm{KH}}$ model.

Based on standard thermodynamics, the Helmholtz free energy $F_{H}$ on the horizon can be defined as
\begin{equation}
   F_{H} = E_{H} - T_{H} S_{H}  .
\label{F_H_def_thermo}
\end{equation}
Assuming the equipartition law of energy on the horizon and substituting $T_{H} S_{H} = E_{H}/2$ given by Eq.\ (\ref{E_ST2_thermo_2}) into Eq.\ (\ref{F_H_def_thermo}) yields 
\begin{equation}
   F_{H} = E_{H} - T_{H} S_{H} = E_{H} - \frac{1}{2}E_{H} = \frac{1}{2}E_{H}  (=  S_{H} T_{H})   .
\label{F_H_2_thermo}
\end{equation}
In addition, from Eq.\ (\ref{F_H_2_thermo}), $dE_{H} - dF_{H}$ is given by
\begin{equation}
  dE_{H} - dF_{H} = dE_{H} - \frac{1}{2}dE_{H} = \frac{1}{2}dE_{H} =dF_{H}    .
\label{dE-dF_equi}
\end{equation}
This equation indicates that $dE_{H} - dF_{H}$ included in Eq.\ (\ref{eq:1stLaw_general_pe_df=0_dEdF_2}) can be replaced by $dF_{H}$, using the equipartition law.

We now apply the equipartition law of energy to the modified thermodynamics relation given by Eq.\ (\ref{eq:1stLaw_general_pe_df=0_dEdF_2}).
Substituting Eq.\ (\ref{dE-dF_equi}) into Eq.\ (\ref{eq:1stLaw_general_pe_df=0_dEdF_2}) yields
\begin{align}
     -dE_{\rm{bulk}}    + W_{e} dV   =  T_{\rm{KH}}  dS_{\rm{BH}}  &=  T_{H} dS_{H}    \notag \\   
                                                                                                                        &=   dE_{H} - dF_{H}   - S_{H} dT_{H}             \notag \\   
                                                                                                                        &=   dF_{H}   - S_{H} dT_{H}       .
\label{eq:1stLaw_general_pe_df=0_dF2}      
\end{align}
The right-hand side of Eq.\ (\ref{eq:1stLaw_general_pe_df=0_dF2}) includes $-S_{H} dT_{H}$.
The right-hand side can be further simplified.
When $T_{H}$ is constant, Eq.\ (\ref{eq:1stLaw_general_pe_df=0_dF2}) is given by
\begin{align}
     -dE_{\rm{bulk}}    + W_{e} dV       &=   dF_{H}      \quad (dT_{H}=0)   .
\label{eq:1stLaw_general_pe_df=0_dF2_Tcst}      
\end{align}
The above equation indicates that the free-energy difference $dF_{H}$ on the cosmological horizon is equivalent to the sum of the negative energy difference ($-dE_{\rm{bulk}}$) and work difference ($W_{e} dV$) in the bulk.
In this sense, a holographic-like connection $E_{\rm{bulk}}=F_{H}$ in a de Sitter universe can be extended to a thermodynamic relation in a constant $T_{H}$ universe.
(The holographic-like connection is summarized in Appendix \ref{Holographic-like connection}.)
The thermodynamic relation is considered to be a kind of extended holographic-like connection.
The free energy on the horizon plays important roles in the thermodynamic relation between the horizon and the bulk.
In addition, multiplying Eq.\ (\ref{eq:Left1stLaw_pe_df=0_drhodV}) by an infinitesimal time $dt$ and coupling the resulting equation with Eq.\ (\ref{eq:1stLaw_general_pe_df=0_dF2_Tcst}) yields
\begin{widetext}
\begin{equation}
     -dE_{\rm{bulk}}    + W_{e} dV        =  \underbrace{- c^{2} V d\rho}_{T_{\rm{GH}}  dS_{\rm{BH}}}  \underbrace{-\- \left ( \frac{\rho c^{2} + p_{e} }{2} \right ) dV}_{\left (\frac{T_{\rm{KH}}}{T_{\rm{GH}}} - 1   \right ) T_{\rm{GH}}  dS_{\rm{BH}} }  
                                                                       = T_{\rm{KH}} dS_{\rm{BH}}  =  T_{H} dS_{H}   = dF_{H}      \quad   (dT_{H}=0)  .   
\label{eq:Left1stLaw_pe_df=0_drhodV_Tcst_dF}      
\end{equation}
\end{widetext}
Equation\ (\ref{eq:Left1stLaw_pe_df=0_drhodV_Tcst_dF}) corresponds to the modified thermodynamic relation in a constant $T_{H}$ universe.
When de Sitter universes are considered (namely, $dV=0$ and $T_{\rm{KH}}=T_{\rm{GH}}$), Eq.\ (\ref{eq:Left1stLaw_pe_df=0_drhodV_Tcst_dF}) reduces to  $- c^{2} V d\rho = T_{\rm{GH}} dS_{\rm{BH}}= dF_{H}$.
From this equation, we can derive the Friedmann equation given by Eq.\ (\ref{eq:Friedmann_drho}).
These results should provide a better understanding of the thermodynamic relation between the horizon and the bulk.

In fact, the holographic entanglement entropy \cite{RyuTakayanagi2006} is equivalent to the formula for the entropy on a cosmological horizon in a de Sitter space \cite{Arias2020}.
In addition, gravity should be related to the relative entropy corresponding to the free-energy difference \cite{Relative_entropy}, based on the holographic entanglement entropy.
The two entropies are usually discussed in a universe at constant temperature.
Accordingly, the free-energy difference $dF_{H}$ included in the modified thermodynamic relation may be related to gravity through the relative entropy and the holographic entanglement entropy. 
These tasks are left for future research.

In this section, we examined the modified thermodynamic relation for constant $f_{\Lambda}(t)$, where an effective pressure $p_{e}$ is considered.
The left-hand side of the relation describes thermodynamic quantities in the bulk and can be interpreted as consisting of contributions from the $\dot{\rho}$ and $\dot{V}$ terms.
The $\dot{\rho}$ term is equivalent to $T_{\rm{GH}}  \dot{S}_{\rm{BH}}$ and the $\dot{V}$ term is equivalent to $[(T_{\rm{KH}}/T_{\rm{GH}}) -1] T_{\rm{GH}}  \dot{S}_{\rm{BH}}$.
The former can lead to the Friedmann equation, while the latter can lead to the acceleration equation by coupling with the derived Friedmann equation.
The magnitude of the $\dot{V}$ term is proportional to the square of the $\dot{\rho}$ term.
In addition, we have applied the equipartition law of energy on the horizon to the right-hand side of the relation, which describes thermodynamic quantities on the horizon.
Consequently, the modified thermodynamic relation is formulated based on the free-energy difference $dF_{H}$ when a constant $T_{\rm{KH}}$ universe is considered.
This thermodynamic relation is considered to be a kind of extended holographic-like connection.
In the next section, we observe typical evolutions of thermodynamic quantities in the modified thermodynamic relation for constant $f_{\Lambda}(t)$, using $\Lambda$CDM models and a constant $T_{\rm{KH}}$ model.

\section{Evolution of thermodynamic quantities in a constant $T_{\rm{KH}}$ model} 
\label{A constant TKH model}

So far, we have not considered specific cosmological models.
In this section, we examine typical evolutions of the thermodynamic quantities in the modified thermodynamic relation for constant $f_{\Lambda}(t)$.
For this, we use $\Lambda$CDM models and a constant $T_{\rm{KH}}$ model.
We first review both models, especially the constant $T_{\rm{KH}}$ model that can describe a constant $T_{\rm{KH}}$ universe.
Then, we observe the evolution of the thermodynamic quantities in both models.
Both models satisfy the modified thermodynamic relation for constant $f_{\Lambda}(t)$.
The background evolution of the universe is then discussed.

The constant $T_{\rm{KH}}$ model is reviewed, according to a previous work \cite{Koma19}.
Systems at constant temperature play important roles in thermodynamics and statistical physics.
In fact, using the constant $T_{\rm{KH}}$ model, we can examine relaxation processes for a universe at constant temperature on a dynamic horizon, as if the dynamic horizon is in contact with a heat bath.
In this sense, the constant $T_{\rm{KH}}$ model should extend the concept of horizons at constant temperature.
The constant $T_{\rm{KH}}$ model should be a good model for studying the relaxation processes for the universe at constant $T_{\rm{KH}}$ \cite{Koma19}.

The constant $T_{\rm{KH}}$ model is obtained from a cosmological model which includes both a power-law term and the equation of state parameter $w$ \cite{Koma19}.
The solution of the power-law model can be applied to the constant $T_{\rm{KH}}$ model.
Therefore, the power-law model is introduced, according to previous works \cite{Koma11,Koma14,Koma19,Koma20}.
In this study, $f_{\Lambda}(t)$ and $h_{\textrm{B}}(t)$ for the power-law model are set to be  
\begin{align}  
f_{\Lambda}(t) =0 \quad \textrm{and} \quad  h_{\textrm{B}}(t) =  \frac{3(1+w)}{2} \Psi_{\alpha} H_{0}^{2}  \left ( \frac{H}{H_{0}} \right )^{\alpha}  ,  
\label{fL_hB__power0}
\end{align}
where $\alpha$ and $\Psi_{\alpha}$ are dimensionless constants whose values are real numbers \cite{Koma11}.
Also, $\alpha$ and $\Psi_{\alpha}$ are independent free parameters, and $\alpha < 2$ and $0 \leq \Psi_{\alpha} \leq 1 $ are considered \cite{Koma19}.
That is, $\Psi_{\alpha}$ is a kind of density parameter for the effective dark energy.
The power-law model considered here corresponds to a pure dissipative universe and satisfies the modified thermodynamic relation for constant $f_{\Lambda}(t)$, because $f_{\Lambda}(t) =0$.
Using the power-law model, we can examine the power-law term systematically.
A similar power-law term for the acceleration equation can be derived from the power-law corrected entropy \cite{Das2008,Radicella2010} and Padmanabhan's holographic equipartition law \cite{Padma2012AB}, as examined in Ref.\ \cite{Koma11}.
(A similar power series was examined in Refs.\ \cite{Valent2015Sola2019,Freaza2002Cardenas2020}.)

In this study, we use Eq.\ (\ref{eq:Back2}), to examine the background evolution of the universe.
Substituting Eq.\ (\ref{fL_hB__power0}) into Eq.\ (\ref{eq:Back2}) yields 
\begin{align}  
    \dot{H}  &=   - \frac{3}{2} (1+ w)  H^{2}  +  \frac{3}{2}   (1+w)  f_{\Lambda}(t)     + h_{\textrm{B}}(t)    \notag \\
                &=   - \frac{3}{2}  (1+ w)  H^{2}  + \frac{3}{2}  (1+ w)  \Psi_{\alpha} H_{0}^{2} \left ( \frac{H}{H_{0}} \right )^{\alpha}  \notag \\
               & = - \frac{3}{2} (1+w) H^{2}  \left (  1 -   \Psi_{\alpha} \left (  \frac{H}{H_{0}} \right )^{\alpha -2} \right )      .  
\label{Modified_eq_power0}
\end{align}
The above equation is equivalent to that for $\Lambda(t)$ models in non-dissipative universes \cite{Koma14,Koma20}.
(In Refs.\ \cite{Koma14,Koma20}, $f_{\Lambda}(t) = \Psi_{\alpha} H_{0}^{2}  ( H / H_{0} )^{\alpha}$ and $h_{\textrm{B}}(t) = 0$ were used for $\Lambda(t)$ models in non-dissipative universes.)
Therefore, we use the solution examined in the previous works.
The solution of Eq.\ (\ref{Modified_eq_power0}) for $\alpha \neq 2$ is written as \cite{Koma14,Koma20}
\begin{equation}  
    \left ( \frac{H}{H_{0}} \right )^{2-\alpha}  =   (1- \Psi_{\alpha})   \tilde{a}^{- \gamma} + \Psi_{\alpha}   ,
\label{eq:Sol_HH0_power0}
\end{equation}
where the normalized scale factor $\tilde{a}$ and the parameter $\gamma$ are given by 
\begin{equation}
\tilde{a} = \frac{a}{a_{0}} \quad \textrm{and} \quad \gamma = \frac{3 (1+w) (2-\alpha)}{2} .
\label{aa0gama_0}
\end{equation}
Here $a_{0}$ is the scale factor at the present time.
Using the power-law model, we can calculate thermodynamic quantities, such as $ T_{\rm{KH}} \dot{S}_{\rm{BH}}$.
The thermodynamic quantities for the power-law model are summarized in Appendix\ \ref{A power-law model} and the results are used in this section.

In fact, a constant $T_{\rm{KH}}$ model is obtained from the power-law model, by setting $w=1/3$ and $\alpha =1$ \cite{Koma19,Koma20}.
Substituting $w=1/3$ and $\alpha =1$ into Eqs. (\ref{fL_hB__power0}), (\ref{Modified_eq_power0}), and (\ref{eq:Sol_HH0_power0}) yields 
\begin{align}  
f_{\Lambda}(t) =0 \quad \textrm{and} \quad  h_{\textrm{B}}(t) =  2 \Psi_{\alpha} H_{0}  H   ,
\label{fL_hB__power_Tcst0}
\end{align}
\begin{align}  
    \dot{H}   &=   - 2  H^{2}  + 2 \Psi_{\alpha} H_{0}  H    ,
\label{Modified_eq_power_Tcst0}
\end{align}
and
\begin{equation}  
    \frac{H}{H_{0}}  =   (1- \Psi_{\alpha})   \tilde{a}^{ -2}  + \Psi_{\alpha}      ,
\label{eq:Sol_HH0_power_Tcst0}
\end{equation}
where $\gamma = 2$ has been used from Eq.\ (\ref{aa0gama_0}).
Equation\ (\ref{Modified_eq_power_Tcst0}) is equivalent to Eq.\ (\ref{Cosmo_T_H_mod_cst}) for a constant $T_{\rm{KH}}$ universe, by replacing $\Psi_{\alpha}$ by $\psi$.
This model corresponds to a constant $T_{\rm{KH}}$ model.
A normalized constant temperature, $T_{\rm{KH}}/T_{\rm{GH},0}=\Psi_{\alpha}$, can be obtained from Eq.\ (\ref{TKH_TGH0_power}), when both $w=1/3$ and $\alpha =1$.
Note that the present model is one viable scenario, in that other cosmological modes can also satisfy Eq.\ (\ref{Cosmo_T_H_mod_cst}), as described in Ref.\ \cite{Koma19}.

In addition, the background evolution of the universe in $\Lambda$CDM models can be examined using the power-law model because Eq.\ (\ref{eq:Sol_HH0_power0}) is equivalent to that for $\Lambda(t)$ models \cite{Koma14,Koma19,Koma20}.
Substituting $w =\alpha =0$ into Eq.\ (\ref{eq:Sol_HH0_power0}) and replacing $\Psi_{\alpha}$ by $\Omega_{\Lambda}$ yields 
\begin{equation}  
 \left ( \frac{H}{H_{0}}  \right )^{2}  =   (1- \Omega_{\Lambda})  \tilde{a}^{ - 3}  + \Omega_{\Lambda}  ,
\label{eq:Sol_HH0_power_LCDM}
\end{equation}
where $\gamma = 3$ is used from Eq.\ (\ref{aa0gama_0}).
Also, $\Omega_{\Lambda}$ is the density parameter for $\Lambda$ and is given by $\Lambda /( 3 H_{0}^{2} ) $ \cite{Koma14}.
The above equation is equivalent to the solution for the $\Lambda$CDM model.
(The influence of radiation is neglected.)
Accordingly, the power-law model for $w =\alpha =0$ is used for the $\Lambda$CDM model.
Of course, the same solution for the $\Lambda$CDM model can be obtained from Eq.\ (\ref{eq:Back2}), when $f_{\Lambda}(t)=\Lambda/3$, $h_{\textrm{B}}(t)=0$, and $w=0$ are used.
The $\Lambda$CDM model satisfies the modified thermodynamic relation for constant $f_{\Lambda}(t)$.

\begin{figure} [t] 
\begin{minipage}{0.495\textwidth}
\begin{center}
\scalebox{0.33}{\includegraphics{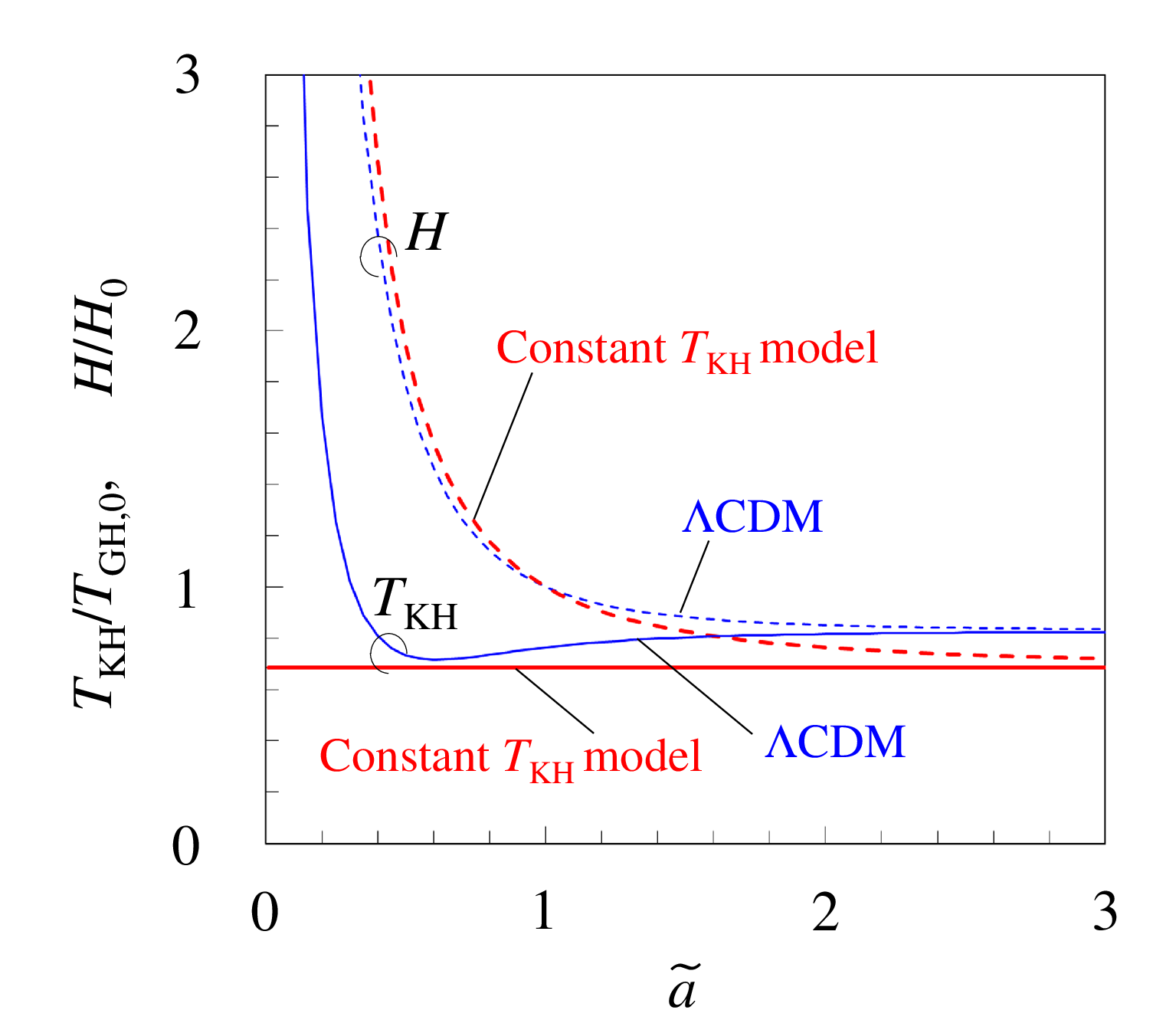}}
\end{center}
\end{minipage}
\caption{Evolution of the normalized Hubble parameter and the normalized Kodama--Hayward temperature for $\Psi_{\alpha}= 0.685$.
The red and blue lines represent the constant $T_{\rm{KH}}$ model and the $\Lambda$CDM model, respectively.
The dashed and solid lines represent the normalized Hubble parameter $H/H_{0}$ and the normalized Kodama--Hayward temperature $T_{\rm{KH}}/T_{\rm{GH},0}$, respectively.
Similar forms of evolution were examined in Refs.\ \cite{Koma19,Koma20}.
Note that $H/H_{0}$ is equivalent to $T_{\rm{GH}}/T_{\rm{GH},0}$, because $T_{\rm{GH}} = \hbar H/(2 \pi  k_{B})$.
 }
\label{Fig-H-a}
\end{figure}

\begin{figure} [t] 
\begin{minipage}{0.495\textwidth}
\begin{center}
\scalebox{0.33}{\includegraphics{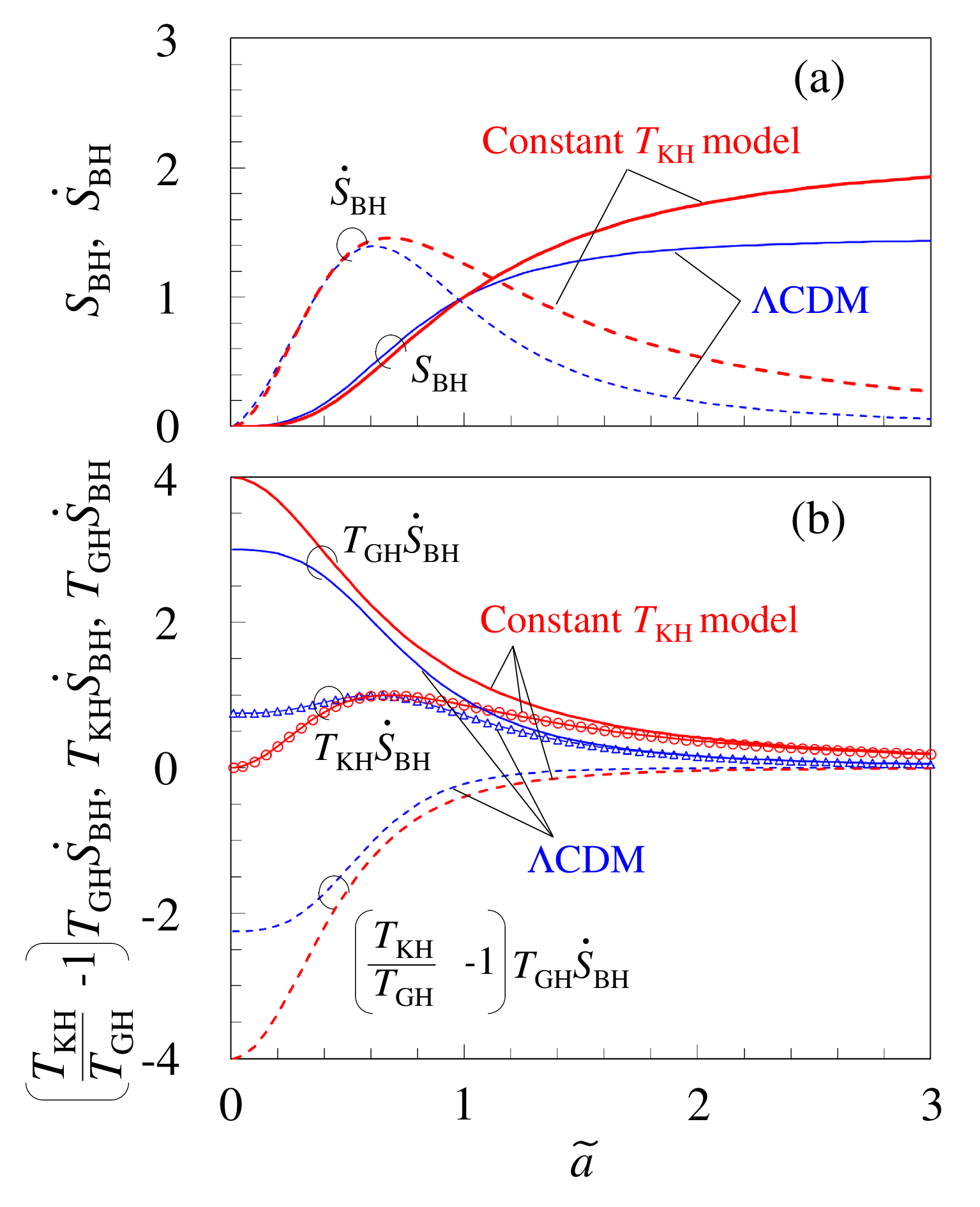}}
\end{center}
\end{minipage}
\caption{Evolution of the normalized thermodynamic quantities for $\Psi_{\alpha}= 0.685$.
(a) Normalized $S_{\rm{BH}}$ and $\dot{S}_{\rm{BH}}$.
(b) Normalized three terms in the modified thermodynamic relation.
The red and blue lines represent the constant $T_{\rm{KH}}$ model and the $\Lambda$CDM model, respectively.
In (a), the solid and dashed lines represent the normalized $S_{\rm{BH}}$ and $\dot{S}_{\rm{BH}}$, respectively.
In (b), the solid and dashed lines represent the normalized $T_{\rm{GH}}  \dot{S}_{\rm{BH}}$ and $[(T_{\rm{KH}}/T_{\rm{GH}}) -1] T_{\rm{GH}}  \dot{S}_{\rm{BH}}$, respectively.
The solid lines with symbols represent the normalized $T_{\rm{KH}}  \dot{S}_{\rm{BH}}$.
For the normalization, see Appendix\ \ref{A power-law model}.
 }
\label{Fig-TdS-a}
\end{figure}

We now observe the evolution of thermodynamic quantities using the $\Lambda$CDM model and the constant $T_{\rm{KH}}$ model.
To examine typical results, $\Psi_{\alpha}$ is set to $0.685$, equivalent to $\Omega_{\Lambda}$ for the $\Lambda$CDM model from the Planck 2018 results \cite{Planck2018}.
Thermodynamic quantities for the power-law model are summarized in Appendix\ \ref{A power-law model} and those results are used for both models.
For the $\Lambda$CDM model, we set $w=\alpha=0$, whereas we set $w=1/3$ and $\alpha =1$ for the constant $T_{\rm{KH}}$ model.
(The thermodynamic quantities depend on the background evolution of the universe.)

Figure\ \ref{Fig-H-a} shows evolutions of $H/H_{0}$ and $T_{\rm{KH}}/T_{\rm{GH},0}$ for the constant $T_{\rm{KH}}$ model and the $\Lambda$CDM model.
The normalized $T_{\rm{KH}}$ for both models is given by Eq.\ (\ref{TKH_TGH0_power}).
The horizontal axis represents the normalized scale factor $\tilde{a}= a/a_{0}$.
Similar forms of evolution have been examined in Refs.\ \cite{Koma19,Koma20}.
As shown in Fig.\ \ref{Fig-H-a}, the normalized $H$ for both models decreases with $\tilde{a}$ and gradually approaches a positive value, corresponding to that for each de Sitter universe. 
In contrast, the normalized $T_{\rm{KH}}$ for the constant $T_{\rm{KH}}$ model is constant during the evolution of the universe even though the Hubble parameter varies with $\tilde{a}$.  
The constant value is given by $T_{\rm{KH}}/T_{\rm{GH},0}=\Psi_{\alpha} =0.685$, from Eq.\ (\ref{TKH_TGH0_power}).
In this way, the normalized $T_{\rm{KH}}$ for the constant $T_{\rm{KH}}$ model is different from that for the $\Lambda$CDM model.
The difference should affect thermodynamic quantities, such as $T_{\rm{KH}}  \dot{S}_{\rm{BH}}$, as examined later.
Of course, finally, the normalized $T_{\rm{KH}}$ approaches $T_{\rm{GH}}/T_{\rm{GH},0}$ for each de Sitter universe.

Figure\ \ref{Fig-TdS-a} shows the evolution of thermodynamic quantities for the constant $T_{\rm{KH}}$ model and the $\Lambda$CDM model.
The normalized $S_{\rm{BH}}$ and $\dot{S}_{\rm{BH}}$ are given by Eqs.\ (\ref{eq:SBH_power}) and (\ref{eq:dSBH_2_3_power_Complicate00}), respectively, and are plotted in Fig.\ \ref{Fig-TdS-a}(a).
As shown in Fig.\ \ref{Fig-TdS-a}(a), the normalized $S_{\rm{BH}}$ for both models increases with $\tilde{a}$ and gradually approaches each positive value.
The normalized $\dot{S}_{\rm{BH}}$ for both models is positive because the normalized $S_{\rm{BH}}$ increases with $\tilde{a}$.
That is, the second law of thermodynamics, $\dot{S}_{\rm{BH}} \geq 0$, is satisfied on the horizon.
Also, the normalized $\dot{S}_{\rm{BH}}$ for both models initially increases and then decreases with $\tilde{a}$, and gradually approaches zero, corresponding to de Sitter universes.
In this way, the evolution of $S_{\rm{BH}}$ and $\dot{S}_{\rm{BH}}$ for the constant $T_{\rm{KH}}$ model is not very different from that for the $\Lambda$CDM model.
We note that similar discussions of the entropic parameters examined above are given in the previous work \cite{Koma19}.

Figure\ \ref{Fig-TdS-a}(b) shows evolutions of three terms included in the modified thermodynamic relation for constant $f_{\Lambda}(t)$.
From Eq.\ (\ref{eq:Left1stLaw_pe_df=0_drhodV}), the relationship between the three terms can be summarized as
\begin{equation}
\underbrace{-\dot{\rho} c^{2} V}_{T_{\rm{GH}}  \dot{S}_{\rm{BH}} }  \underbrace{-\- \left ( \frac{\rho c^{2} + p_{e} }{2} \right )  \dot{V}}_{\left (\frac{T_{\rm{KH}}}{T_{\rm{GH}}} - 1   \right ) T_{\rm{GH}}  \dot{S}_{\rm{BH}} }  
= T_{\rm{KH}}  \dot{S}_{\rm{BH}}    .
\label{eq:Left1stLaw_pe_df=0_drhodV_3terms}      
\end{equation}
First, we observe the evolution of the right-hand side of Eq.\ (\ref{eq:Left1stLaw_pe_df=0_drhodV_3terms}), namely $T_{\rm{KH}}  \dot{S}_{\rm{BH}}$.
The normalized $T_{\rm{KH}}  \dot{S}_{\rm{BH}}$ for both models is given by Eq.\ (\ref{TKHdSBH_power_Norm}).
As shown in Fig.\ \ref{Fig-TdS-a}(b), initially, the normalized $T_{\rm{KH}}  \dot{S}_{\rm{BH}}$ for the constant $T_{\rm{KH}}$ model is different from that for the $\Lambda$CDM model.
This is because the normalized $T_{\rm{KH}}$ for both models is different from each other, especially in the initial stage, as examined in Fig.\ \ref{Fig-H-a}.
Note that $T_{\rm{KH}}  \dot{S}_{\rm{BH}}$ for the constant $T_{\rm{KH}}$ model is equivalent to $\dot{F}_{H}$, because of $T_{\rm{KH}} dS_{\rm{BH}}=dF_{H}$ given by Eq.\ (\ref{eq:Left1stLaw_pe_df=0_drhodV_Tcst_dF}).

Next, we observe evolutions of the first and second terms on the left-hand side of Eq.\ (\ref{eq:Left1stLaw_pe_df=0_drhodV_3terms}).
The first and second terms, namely the $\dot{\rho}$ and $\dot{V}$ terms, are equivalent to $T_{\rm{GH}}  \dot{S}_{\rm{BH}}$ and $[(T_{\rm{KH}}/T_{\rm{GH}}) -1] T_{\rm{GH}}  \dot{S}_{\rm{BH}}$, respectively.
The two normalized thermodynamic quantities for both models are given by Eqs.\ (\ref{eq:TGH-dSBH_2_power_Norm}) and (\ref{eq:2nd-term_power_Norm}) and are plotted in Fig.\ \ref{Fig-TdS-a}(b).
As shown in Fig.\ \ref{Fig-TdS-a}(b), initially, the evolutions of the two normalized terms for the constant $T_{\rm{KH}}$ model are quantitatively different from those for the $\Lambda$CDM model.
The normalized $T_{\rm{GH}}  \dot{S}_{\rm{BH}}$ is positive, whereas the normalized $[(T_{\rm{KH}}/T_{\rm{GH}}) -1] T_{\rm{GH}}  \dot{S}_{\rm{BH}}$ is negative.
Specifically, the two initial values for the constant $T_{\rm{KH}}$ model are $4$ and $-4$, respectively, whereas the two initial values for the $\Lambda$CDM model are $3$ and $-9/4$, respectively, as examined in Appendix \ref{A power-law model}.
The sum of the normalized first and second terms is equivalent to the normalized $T_{\rm{KH}}  \dot{S}_{\rm{BH}}$, because this relation is given by Eq.\ (\ref{eq:Left1stLaw_pe_df=0_drhodV_3terms}).
(The initial value depends on $w$, where $w=0$ and $w=1/3$ are set for the $\Lambda$CDM and constant $T_{\rm{KH}}$ models, respectively, as examined in Appendix \ref{A power-law model}.)

\begin{figure} [t] 
\begin{minipage}{0.495\textwidth}
\begin{center}
\scalebox{0.33}{\includegraphics{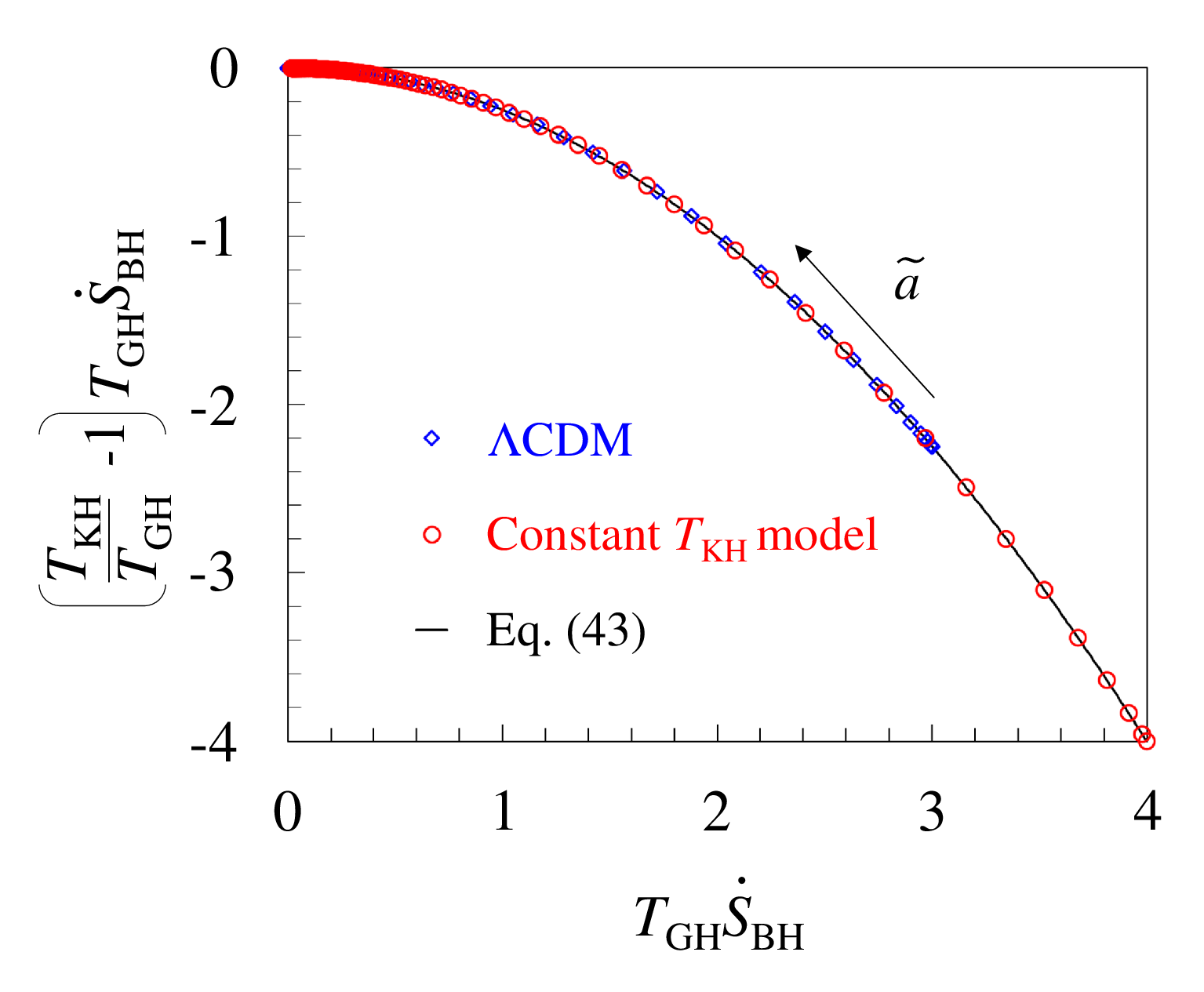}}
\end{center}
\end{minipage}
\caption{Relationship between two normalized thermodynamic quantities during the evolution of the universe for $\Psi_{\alpha}= 0.685$.
The quantities for the horizontal and vertical axes correspond to the $\dot{\rho}$ and $\dot{V}$ terms on the left-hand side of Eq.\ (\ref{eq:Left1stLaw_pe_df=0_drhodV_3terms}), respectively.
The red circles and blue diamonds represent the constant $T_{\rm{KH}}$ model and $\Lambda$CDM model, respectively.
The arrow indicates the direction in which $\tilde{a}$ increases from $0$ to $10$, with increments of $0.05$.
The initial coordinate values corresponding to $\tilde{a}=0$ for the constant $T_{\rm{KH}}$ model and the $\Lambda$CDM model are $(4, -4)$ and $(3, -9/4)$, respectively.
For the initial values, see Fig.\ \ref{Fig-TdS-a}(b) and the text.
 }
\label{Fig-(TKHTGH-1)dS-TGHdS}
\end{figure}

As discussed previously, Eqs.\ (\ref{eq:1st-2nd-terms}) and (\ref{eq:1st-2nd-terms_Norm}) indicate that the magnitude of the second term (the $\dot{V}$ term) is proportional to the square of the first term (the $\dot{\rho}$ term).
To confirm this, we examine the relationship between the two normalized terms for the constant $T_{\rm{KH}}$ model and the $\Lambda$CDM model, as shown in Fig.\ \ref{Fig-(TKHTGH-1)dS-TGHdS}.
The arrow indicates the direction in which the normalized scale factor $\tilde{a}$ increases from $0$ to $10$.
In this figure, the initial coordinate values corresponding to $\tilde{a}=0$ for the constant $T_{\rm{KH}}$ model and the $\Lambda$CDM model are $(4, -4)$ and $(3, -9/4)$, respectively.
As $\tilde{a}$ increases, the normalized plots gradually approach the origin of the coordinates, $(0, 0)$, corresponding to those for de Sitter universes.
We can confirm that all the normalized plots for both models show a quadratic curve. 
The quadratic curve is described by Eq.\ (\ref{eq:1st-2nd-terms_Norm}), which is derived without using specific cosmological models.
The relationship between the two terms is universal when the modified thermodynamic relation for constant $f_{\Lambda}(t)$ is satisfied.

In this section, we observed typical evolutions of the thermodynamic quantities in the modified thermodynamic relation using the $\Lambda$CDM model and the constant $T_{\rm{KH}}$ model.
The constant $T_{\rm{KH}}$ model is simply one viable scenario with a constant horizon temperature \cite{Koma19}.
However, the results for the constant $T_{\rm{KH}}$ model will contribute to the study of thermodynamics and statistical physics on dynamic horizons because the horizon temperature is constant.
For example, a constant $T_{\rm{KH}}$ universe always satisfies the holographic equipartition law of energy.
Therefore, the emergence of cosmic space based on the law \cite{Padma2010,Verlinde1,HDE,Padmanabhan2004,ShuGong2011,Koma14,Koma15,Koma16,Koma17,Koma19,Koma20,Padma2012AB,Cai2012,Hashemi,Moradpour,Wang,Koma10,Koma11,Koma12,Koma18,Krishna20172019,Mathew2022,Chen2022,Luciano,Mathew2023,Mathew2023b,Pad2017,Tu2018,Tu2019,Chen2024}, 
which can lead to cosmological equations, should be discussed from a different viewpoint, using the modified thermodynamic relation.
Of course, in general, the modified thermodynamic relation is given by Eq.\ (\ref{eq:1stLaw_general}) and the cosmological equations are given by Eqs. (\ref{eq:General_FRW01})--(\ref{eq:drho_General}).
In this case, we should examine the cosmological constant problem, by applying the second law of thermodynamics to the equations and extending a method used in previous works \cite{Koma11,Koma12}.
These tasks are left for future research.

The modified thermodynamic relation examined in this paper should help to understand the properties of various cosmological models from a thermodynamic viewpoint.
The present study should provide new insights into the discussion of thermodynamic cosmological scenarios.

\section{Conclusions}
\label{Conclusions}

We examined the holographic-like thermodynamic relation between a cosmological horizon and the bulk by applying a general formulation for cosmological equations to the first law of thermodynamics.
For the general formulation, both an effective pressure $p_{e}$ for dissipative universes and an extra driving term $f_{\Lambda}(t)$ for non-dissipative universes are phenomenologically assumed in a flat FLRW universe.
We derived the modified thermodynamic relation that includes both $p_{e}$ and an additional time-derivative term $\dot{f}_{\Lambda}(t)$, by applying the general formulation to the first law.
When $f_{\Lambda}(t)$ is constant, the modified thermodynamic relation reduces to the formulation of the first law in standard cosmology. 

Next, we examined the modified thermodynamic relation for constant $f_{\Lambda}(t)$, with $p_{e}$ considered.
The left-hand side of the relation describes thermodynamic quantities in the bulk and can be interpreted as consisting of contributions from two terms, namely the $\dot{\rho}$ and $\dot{V}$ terms.
It is found that the $\dot{\rho}$ term is equivalent to $T_{\rm{GH}}  \dot{S}_{\rm{BH}}$, and the $\dot{V}$ term is equivalent to $[(T_{\rm{KH}}/T_{\rm{GH}}) -1] T_{\rm{GH}}  \dot{S}_{\rm{BH}}$.
The former equivalence can lead to the Friedmann equation,
while the latter can lead to the acceleration equation by coupling with the derived Friedmann equation.
The magnitude of the $\dot{V}$ term is proportional to the square of the $\dot{\rho}$ term.
In addition, we applied the equipartition law of energy on the horizon to the $T_{\rm{KH}}  \dot{S}_{\rm{BH}}$ term, namely the right-hand side of the modified thermodynamic relation.
Consequently, when $T_{\rm{KH}}$ is constant, the modified thermodynamic relation reduces to $-dE_{\rm{bulk}}    + W_{e} dV  =  dF_{H}$.
This thermodynamic relation is considered to be a kind of extended holographic-like connection.

Finally, we observed typical evolutions of the thermodynamic quantities in the modified thermodynamic relation for constant $f_{\Lambda}(t)$ using $\Lambda$CDM models and a constant $T_{\rm{KH}}$ model.
Initially, thermodynamic quantities which include $T_{\rm{KH}}$ for both models are different although the evolution of the Hubble parameter is similar between the two models.
These thermodynamic quantities for both models gradually approach constant values corresponding to those for de Sitter universes.

The assumptions used in this paper have not yet been established but are considered to be viable scenarios.
The modified thermodynamic relation for a constant $T_{\rm{KH}}$ universe implies that the free-energy difference $dF_{H}$ on the horizon plays important roles.
The present study should contribute to a better understanding of the thermodynamic relation between the cosmological horizon and the bulk and should provide new insights into thermodynamics and cosmological equations.

\appendix

\section{Holographic-like connection}
\label{Holographic-like connection}

Padmanabhan \cite{Pad2017} derived an energy-balance relation $\rho c^2 V = T_{\rm{GH}} S_{\rm{BH}}$,
which is essentially equivalent to a holographic-like connection, $E_{\rm{bulk}}=F_{H}$. 
This appendix briefly reviews the holographic-like connection based on previous works \cite{Koma18,Koma20}.

We first calculate $E_{\rm{bulk}}$ from the Friedmann equation in standard cosmology \cite{Koma18,Koma20}.
Substituting $f_{\Lambda}(t) =0$ into Eq.\ (\ref{eq:General_FRW01}), the Friedmann equation is written as  
\begin{equation}
 H^2      =  \frac{ 8\pi G }{ 3 } \rho              .                                                 
\label{eq:FRW01} 
\end{equation} 
Substituting Eqs.\ (\ref{eq:FRW01}) and (\ref{eq:V}) into Eq.\ (\ref{Ebulk}) yields 
\begin{align}
E_{\rm{bulk}}  &= \rho c^{2} V =  \frac{ 3 H^{2} c^{2} }{ 8 \pi G }      \frac{4}{3} \pi \left ( \frac{c}{H} \right )^3   = \frac{1}{2}   \frac{ c^{5} }{ G }  \left ( \frac{1}{H} \right )       .
\label{Ebulk_FRW}
\end{align}
Next, we calculate $F_{H}$ from the equipartition law of energy on the horizon, using $S_{H} = S_{\rm{BH}}$ and $T_{H} = T_{\rm{GH}}$,
where $T_{\rm{GH}}$ is the Gibbons--Hawking temperature.
Substituting Eqs.\ (\ref{eq:SBH2}) and (\ref{eq:T_GH}) into Eq.\ (\ref{F_H_2_thermo}) yields 
\begin{align}
   F_{H}                    =  S_{H} T_{H}  =  S_{\rm{BH}} T_{\rm{GH}} &= \left ( \frac{ \pi k_{B} c^5 }{ \hbar G } \right )  \frac{1}{H^2}    \left (\frac{ \hbar H}{   2 \pi  k_{B}  } \right )   \notag \\
                                                                                        &= \frac{1}{2}   \frac{ c^{5} }{ G }  \left ( \frac{1}{H} \right )  .
\label{F_H_SBH_TGH}
\end{align}
From Eqs.\ (\ref{Ebulk_FRW}) and (\ref{F_H_SBH_TGH}), we have the relation 
\begin{equation}
E_{\rm{bulk}}  =  F_{H}       .
\label{Ebulk_FH}
\end{equation}
This consistency is the holographic-like connection in standard cosmology \cite{Koma18,Koma20}.
(Note that only the magnitudes of $E_{\rm{bulk}}$ and $F_{H}$ are considered.)
The holographic-like connection has not yet been established. 
In fact, the energy in the equipartition law has been discussed from different viewpoints, e.g., in the works of Verlinde \cite{Verlinde1} and Padmanabhan \cite{Padma2010,Padma2012AB}, as described in Ref.\ \cite{Koma20}.

In the above derivation, standard cosmology and the equipartition law of energy on the horizon are assumed.
In addition, $T_{\rm{GH}}$ is used for $T_{H}$, instead of a dynamical Kodama--Hayward temperature $T_{\rm{KH}}$. 
In this sense, the holographic-like connection should correspond to a thermodynamic relation for de Sitter universes whose Hubble volume is constant.
In the present paper, we extend this connection to the thermodynamic relation in universes whose Hubble volume varies with time.

\section{Thermodynamic quantities for a power-law model} 
\label{A power-law model}

In this appendix, we calculate the thermodynamic quantities for a power-law model \cite{Koma11,Koma16,Koma19}.
In fact, a constant $T_{\rm{KH}}$ model is obtained from the power-law model by setting $w=1/3$ and $\alpha =1$.
The thermodynamic quantities calculate here are used for the constant $T_{\rm{KH}}$ model. 
($\Lambda$CDM models are descried in Sec.\ \ref{A constant TKH model}.)

First, we again focus on the power-law model.
From Eq.\ (\ref{Modified_eq_power0}), the cosmological equation is given by
\begin{align}  
    \dot{H}      & = - \frac{3}{2} (1+w) H^{2}  \left (  1 -   \Psi_{\alpha} \left (  \frac{H}{H_{0}} \right )^{\alpha -2} \right )      ,  
\label{Modified_eq_power}
\end{align}
where $\alpha < 2$ and $0 \leq \Psi_{\alpha} \leq 1 $ are considered \cite{Koma19}.
From Eq.\ (\ref{eq:Sol_HH0_power0}), the solution of Eq.\ (\ref{Modified_eq_power}) for $\alpha \neq 2$ is written as \cite{Koma14,Koma20}
\begin{equation}  
    \left ( \frac{H}{H_{0}} \right )^{2-\alpha}  =   (1- \Psi_{\alpha})   \tilde{a}^{- \gamma} + \Psi_{\alpha}   ,
\label{eq:Sol_HH0_power}
\end{equation}
where $\tilde{a}$ and $\gamma$ are given by 
\begin{equation}
\tilde{a} = \frac{a}{a_{0}} \quad \textrm{and} \quad \gamma = \frac{3 (1+w) (2-\alpha)}{2} .
\label{aa0gama}
\end{equation}

We now calculate several thermodynamic quantities for the power-law model, according to Ref.\ \cite{Koma19}.
From Eq.\ (\ref{eq:SBH2}), the normalized $S_{\rm{BH}}$ is written as
\begin{equation}
\frac{S_{\rm{BH}}}{ S_{\rm{BH},0} } =  \left ( \frac{ H }{ H_{0} } \right )^{-2}  .
\label{eq:SBHSBH0_0a}      
\end{equation}
Substituting Eq.\ (\ref{eq:Sol_HH0_power}) into Eq.\ (\ref{eq:SBHSBH0_0a}) yields  
\begin{align}  
\frac{S_{\rm{BH}}}{ S_{\rm{BH},0} } &=    \left [     (1- \Psi_{\alpha})  \tilde{a}^{ - \gamma  }  + \Psi_{\alpha}     \right ]^{\frac{2}{\alpha-2}}      .
\label{eq:SBH_power}      
\end{align}  
In addition, we calculate $\dot{S}_{\rm{BH}}$ for the power-law model.
Substituting Eq.\ (\ref{Modified_eq_power}) into $\dot{S}_{\rm{BH}}$ given by Eq.\ (\ref{eq:dSBH}) and applying Eq.\ (\ref{eq:Sol_HH0_power}) yields \cite{Koma19}
\begin{align}  
\dot{S}_{\rm{BH}}   
&=  \frac{-2K \dot{H} }{H^{3}}  =  \frac{2K}{H_{0}}  \left ( \frac{- \dot{H} }{H^{2}} \right )    \frac{H_{0}}{H}                                    \notag \\
&=   \frac{3K}{H_{0}} \frac{  (1+w) (1- \Psi_{\alpha})  \tilde{a}^{- \gamma}    }{      \left [ (1- \Psi_{\alpha})  \tilde{a}^{- \gamma} + \Psi_{\alpha}  \right ]^{ \frac{3-\alpha}{2- \alpha} }  } .
\label{eq:dSBH_2_3_power_Complicate}      
\end{align}   
Using Eq.\ (\ref{eq:dSBH_2_3_power_Complicate}) and $S_{\rm{BH},0}= K/H_{0}^{2}$, the normalized $\dot{S}_{\rm{BH}}$ is written as \cite{Koma19}
\begin{align}  
\frac{\dot{S}_{\rm{BH}}}{ S_{\rm{BH},0} H_{0}}  &=  \frac{  3 (1+w) (1- \Psi_{\alpha})   \tilde{a}^{- \gamma}     }{   \left [ (1- \Psi_{\alpha})   \tilde{a}^{- \gamma}   + \Psi_{\alpha}  \right ]^{ \frac{3- \alpha}{2- \alpha} }  } .
\label{eq:dSBH_2_3_power_Complicate00}      
\end{align}  
This equation indicates that $\dot{S}_{\rm{BH}} \geq 0$ is satisfied when $w \ge -1$ and $0 \le \Psi_{\alpha} \le 1$ are considered.
The second law of thermodynamics and the maximization of the entropy have been examined in previous works \cite{Koma14,Koma19}.

Next, we calculate the normalized Kodama--Hayward temperature for the power-law model.
From Eq.\ (\ref{eq:T_KH}), the normalized Kodama--Hayward temperature is written as \cite{Koma19,Koma20}
\begin{equation}
\frac{T_{\rm{KH}}}{ T_{\rm{GH},0} } =  \frac{H}{H_{0}}  \left ( 1 + \frac{ \dot{H} }{ 2 H^{2} }\right )   ,
\label{eq:TKH_TGH0}      
\end{equation}
where $T_{\rm{GH},0} =  \hbar H_{0} / (2 \pi  k_{B})$ is used.
From Eq.\ (\ref{Modified_eq_power}), the power-law model satisfies $1 + \frac{ \dot{H} }{ 2 H^{2} } \ge 0$, because $w \le 1/3$ and the non-negative driving terms are considered.
Therefore, the normalized $T_{\rm{KH}}$ is non-negative in an expanding universe.
Substituting Eq.\ (\ref{Modified_eq_power}) into Eq.\ (\ref{eq:TKH_TGH0}), substituting Eq.\ (\ref{eq:Sol_HH0_power}) into the resulting equation, and performing several calculations yields \cite{Koma19}
\begin{align}
\frac{T_{\rm{KH}}}{ T_{\rm{GH},0} }    &=  \frac{(1-3w) (1- \Psi_{\alpha})  \tilde{a}^{ - \gamma}    + 4 \Psi_{\alpha}   }{4  \left [ (1- \Psi_{\alpha})  \tilde{a}^{ - \gamma}  + \Psi_{\alpha}  \right ]^{\frac{1-\alpha}{2-\alpha}} } .   
\label{TKH_TGH0_power}
\end{align}
When $w = 1/3$ and $\alpha =1$ are considered, this equation reduces to a constant value given by $T_{\rm{KH}}/T_{\rm{GH},0}=\Psi_{\alpha}$.
That is, the power-law model for $w = 1/3$ and $\alpha =1$ corresponds to a constant $T_{\rm{KH}}$ model \cite{Koma19}.

Using Eqs.\ (\ref{TKH_TGH0_power}) and (\ref{eq:dSBH_2_3_power_Complicate00}), the normalized $T_{\rm{KH}} \dot{S}_{\rm{BH}}$ is given by
\begin{align}
\frac{T_{\rm{KH}}  \dot{S}_{\rm{BH}} }{ T_{\rm{GH},0}   (S_{\rm{BH},0} H_{0})     }     &=  \frac{(1-3w) (1- \Psi_{\alpha})  \tilde{a}^{ - \gamma}    + 4 \Psi_{\alpha}   }{4  \left [ (1- \Psi_{\alpha})  \tilde{a}^{ - \gamma}  + \Psi_{\alpha}  \right ]^{2} }  \notag \\
                                                                                                                     & \times 3 (1+w) (1- \Psi_{\alpha})   \tilde{a}^{- \gamma}    .    
\label{TKHdSBH_power_Norm}
\end{align}
The power-law model satisfies $T_{\rm{KH}} \dot{S}_{\rm{BH}} \ge 0$.

Finally, we calculate the first and second terms on the left-hand side of Eq.\ (\ref{eq:Left1stLaw_pe_df=0_drhodV_3terms}), namely the $\dot{\rho}$ and $\dot{V}$ terms, equivalent to $T_{\rm{GH}}  \dot{S}_{\rm{BH}}$ and $[(T_{\rm{KH}}/T_{\rm{GH}}) -1] T_{\rm{GH}}  \dot{S}_{\rm{BH}}$, respectively.
Substituting Eq.\ (\ref{Modified_eq_power}) into Eq.\ (\ref{eq:TGH-dSBH_2}) and applying Eq.\ (\ref{eq:Sol_HH0_power}) yields
\begin{align}
T_{\rm{GH}}  \dot{S}_{\rm{BH}}    &=  \left ( \frac{c^5 }{G} \right ) \left ( \frac{ - \dot{H} }{H^{2}} \right )   \notag \\      
                                               &=  \left ( \frac{c^5 }{G} \right )   \frac{3}{2} (1+w)   \left (  1 -   \Psi_{\alpha} \left (  \frac{H}{H_{0}} \right )^{\alpha -2} \right )      \notag \\        
                                               &= \left ( \frac{c^5 }{G} \right )   \frac{3}{2} (1+w)   \left (  1 -   \frac{\Psi_{\alpha}}{ (1- \Psi_{\alpha})   \tilde{a}^{- \gamma}   + \Psi_{\alpha} } \right )      \notag \\        
                                               &=  \left ( \frac{c^5 }{G} \right )   \frac{ \frac{3}{2} (1+w)  (1- \Psi_{\alpha})   \tilde{a}^{- \gamma} }{ (1- \Psi_{\alpha})   \tilde{a}^{- \gamma}   + \Psi_{\alpha} }  .
\label{eq:TGH-dSBH_2_power}      
\end{align}
Similarly, substituting Eq.\ (\ref{Modified_eq_power}) into Eq.\ (\ref{eq:2nd-term}) and applying Eq.\ (\ref{eq:Sol_HH0_power}) yields
\begin{align}
     & \left (\frac{T_{\rm{KH}}}{T_{\rm{GH}}} - 1   \right ) T_{\rm{GH}}  \dot{S}_{\rm{BH}}   = \left ( \frac{  \dot{H} }{ 2H^{2}} \right )  \left ( \frac{c^5 }{G} \right ) \left ( \frac{ - \dot{H} }{H^{2}} \right )       \notag \\   
                                                                                                                             &=   - \frac{1}{2}  \left ( \frac{c^5 }{G} \right )  \left [  \frac{ \frac{3}{2} (1+w)  (1- \Psi_{\alpha})   \tilde{a}^{- \gamma} }{ (1- \Psi_{\alpha})   \tilde{a}^{- \gamma}   + \Psi_{\alpha} }  \right ]^{2}    .
\label{eq:2nd-term_power}      
\end{align}
The same equation can be obtained from Eqs.\ (\ref{eq:1st-2nd-terms}) and (\ref{eq:TGH-dSBH_2_power}).
For the normalization, dividing Eq.\ (\ref{eq:TGH-dSBH_2_power}) by $T_{\rm{GH},0} (S_{\rm{BH},0} H_{0})$ yields the normalized first term:
\begin{align}  
  \frac{ T_{\rm{GH}}  \dot{S}_{\rm{BH}} }{ T_{\rm{GH},0}  (S_{\rm{BH},0} H_{0}) }  &=  \frac{  3 (1+w) (1- \Psi_{\alpha})   \tilde{a}^{- \gamma}     }{ (1- \Psi_{\alpha})   \tilde{a}^{- \gamma}   + \Psi_{\alpha}  } .
\label{eq:TGH-dSBH_2_power_Norm}      
\end{align}  
To calculate the initial value, we set $\tilde{a}=0$, although inflation of the early universe is not discussed.
When $\tilde{a}=0$, the normalized first term reduces to $3(1+w)$ and, therefore, the normalized initial value for $w=1/3$ and $w=0$ is $4$ and $3$, respectively.
In this way, the normalized initial value depends on $w$.
Similarly, from Eq.\ (\ref{eq:2nd-term_power}), the normalized second term is given by
\begin{align}
   \frac{ \left (\frac{T_{\rm{KH}}}{T_{\rm{GH}}} - 1   \right ) T_{\rm{GH}}  \dot{S}_{\rm{BH}} }{ T_{\rm{GH},0}  (S_{\rm{BH},0} H_{0}) }   &=   - \frac{1}{4}   \left [  \frac{ 3 (1+w)  (1- \Psi_{\alpha})   \tilde{a}^{- \gamma} }{ (1- \Psi_{\alpha})   \tilde{a}^{- \gamma}   + \Psi_{\alpha} }  \right ]^{2}    .
\label{eq:2nd-term_power_Norm}      
\end{align}
When $\tilde{a}=0$, the normalized second term for $w=1/3$ and $w=0$ is $-4$ and $-9/4$, respectively.

Note that the numerical coefficients in Eqs. (\ref{eq:TGH-dSBH_2_power_Norm}) and (\ref{eq:2nd-term_power_Norm}) are different from those in Eqs. (\ref{eq:TGH-dSBH_2_power}) and (\ref{eq:2nd-term_power}), respectively,
because Eqs.\ (\ref{eq:TGH-dSBH_2_power_Norm}) and (\ref{eq:2nd-term_power_Norm}) have been normalized by $T_{\rm{GH},0} =  \hbar H_{0} / (2 \pi  k_{B})$, which includes a coefficient of $1/2$.

Substituting Eq.\ (\ref{eq:TGH-dSBH_2_power_Norm}) into Eq.\ (\ref{eq:2nd-term_power_Norm}) yields 
\begin{align}
   \frac{ \left (\frac{T_{\rm{KH}}}{T_{\rm{GH}}} - 1   \right ) T_{\rm{GH}}  \dot{S}_{\rm{BH}} }{ T_{\rm{GH},0}  (S_{\rm{BH},0} H_{0}) }   &=   - \frac{1}{4}   \left [   \frac{ T_{\rm{GH}}  \dot{S}_{\rm{BH}} }{ T_{\rm{GH},0}  (S_{\rm{BH},0} H_{0}) } \right ]^{2}    .
\label{eq:1st-2nd-terms_power_Norm}      
\end{align}
This equation is equivalent to Eq.\ (\ref{eq:1st-2nd-terms_Norm}), which is derived without using specific cosmological models.
Accordingly, Eq.\ (\ref{eq:1st-2nd-terms_power_Norm}) is a universal relationship between the normalized first and second terms.

The power-law model is reduced to a constant $T_{\rm{KH}}$ model by setting $w=1/3$ and $\alpha =1$ \cite{Koma19,Koma20}.
Therefore, the thermodynamic quantities for the power-law model can be applied to the constant $T_{\rm{KH}}$ model.
In Sec.\ \ref{A constant TKH model}, we discuss the constant $T_{\rm{KH}}$ model, using the power-law model for $w=1/3$ and $\alpha =1$.

\section{Euclidean action and a general `action--entropy relation'} 
\label{A consistency of the equipartition law}

The equipartition law of energy on the horizon used in Sec.\ \ref{Equipartition law} has not yet been established but is considered to be a viable scenario.
In this appendix, we discuss the equipartition law from a different viewpoint.
In fact, $E_{H} = 2S_{H}T_{H}$ given by Eq.\ (\ref{E_ST2_thermo_2}), which is based on the equipartition law, can be obtained from Euclidean action and a general `action--entropy relation'.
This derivation is examined.

We first introduce a general relationship between an action $\mathcal{A}$ and an entropy $S$, according to the work of de Broglie \cite{Broglie}.
The general `action--entropy relation' can be written as \cite{Broglie}
\begin{align}
   \frac{\mathcal{A}}{\hbar} = \frac{S}{k_{B}} ,
\label{eq:de-broglie_Action}      
\end{align}
where the Planck constant $h$ has been replaced by the reduced Planck constant $\hbar$ \cite{ActionEntropy}.
We assume that Eq.\ (\ref{eq:de-broglie_Action}) is satisfied on a cosmological horizon.

Next, according to the work of Gibbons and Hawking \cite{GibbonsHawking1977Action}, we introduce the relationship between the Helmholtz free energy $F$ and Euclidean action $I$ of black holes, which is approximately written as \cite{York1988,Whiting1990,Wei2022}
\begin{align}
   I \approx \hbar \beta F ,
\label{eq:Euclidean action}      
\end{align}
where the Wick rotation ($ t \rightarrow - i \hbar \beta $) has been used and $\beta$ represents the inverse temperature given by $1/(k_{B}T)$.
In fact, Eq.\ (\ref{eq:Euclidean action}) can be applied to the cosmological horizon such as the Hubble horizon, as examined in the work of Arias \textit{et al}. \cite{Arias2020}.
Therefore, we assume that Eq.\ (\ref{eq:Euclidean action}) can be applied to the Hubble horizon.

Accepting these assumptions, setting $\mathcal{A}=I$, and substituting Eq.\ (\ref{eq:Euclidean action}) into Eq.\ (\ref{eq:de-broglie_Action}) yields
\begin{align}
   \frac{S_{H}}{k_{B}} = \frac{\mathcal{A}} {\hbar} = \frac{I}{\hbar} \approx \frac{\hbar \beta F_{H}}{\hbar} = \frac{F_{H}}{k_{B}T_{H}}, 
\label{eq:Action_F}      
\end{align}
where $S$, $T$, and $F$ have been replaced by $S_{H}$, $T_{H}$, and $F_{H}$, respectively.
$T_{\rm{KH}}$ is not used.
Consequently, the free energy $F_{H}$ on the horizon is given by 
\begin{align}
 F_{H} \approx T_{H} S_{H}  .
\label{eq:F_TS_Action}      
\end{align}
This equation is consistent with Eq.\ (\ref{F_H_2_thermo}).
Substituting Eq.\ (\ref{eq:F_TS_Action}) into the definition of the free energy, $F_{H}=E_{H}-T_{H}S_{H}$, yields
\begin{align}
  E_{H} \approx 2S_{H}T_{H}  ,
\label{eq:E_2ST_Action}      
\end{align}
which is consistent with Eq.\ (\ref{E_ST2_thermo_2}). 
In addition, from $E_{H} \approx 2S_{H}T_{H}$, we can obtain $E_{H} \approx  N_{\rm{sur}} (k_{B} T_{H} /2)$, corresponding to Eq.\ (\ref{E_equip}), when $N_{\rm{sur}} = 4 S_{H} /k_{B}$ given by Eq.\ (\ref{N_sur}) is applied.
In this way, the equipartition law is likely consistent with the free energy calculated from Euclidean action and the general `action--entropy relation'.
This consistency may be satisfied not only in de Sitter universes but also in a constant $T_{\rm{KH}}$ universe discussed in the present study.
The above result implies that the equipartition law is a viable scenario.
Several assumptions used here have not yet been established and further studies are needed.


\begin{thebibliography}{99}

\bibitem{PERL1998_Riess1998} S. Perlmutter \textit{et al.},  Nature (London) \textbf{391}, 51 (1998); A. G. Riess \textit{et al.}, Astron. J. \textbf{116}, 1009 (1998).
\bibitem{Hubble2017} O. Farooq, F. R. Madiyar, S. Crandall, B. Ratra, Astrophys. J. \textbf{835}, 26 (2017).     
\bibitem{Planck2018} N. Aghanim \textit{et al.}, Astron. Astrophys. \textbf{641}, A6 (2020).        




\bibitem{FreeseOverduin}
K. Freese, F. C. Adams, J. A. Frieman, E. Mottola, Nucl. Phys. \textbf{B287}, 797 (1987); 
J. M. Overduin, F. I. Cooperstock, Phys. Rev. D \textbf{58}, 043506 (1998).
\bibitem{Nojiri2006etc}  
S. Nojiri, S. D. Odintsov, Phys. Lett. B \textbf{639}, 144 (2006);
J. Sol\`{a}, J. Phys. Conf. Ser. \textbf{453},  012015 (2013).
\bibitem{Sola_2009-2022} 
S. Basilakos, M. Plionis, J. Sol\`{a}, Phys. Rev. D \textbf{80}, 083511 (2009);
M. Rezaei, J. Sol\`{a} Peracaula, M. Malekjani, Mon. Not. R. Astron. Soc. \textbf{509}, 2593 (2022).
%
\bibitem{Valent2015Sola2019}  
A. G\'{o}mez-Valent, J. Sol\`{a}, S. Basilakos, J. Cosmol. Astropart. Phys. 01 (2015) 004;
M. Rezaei, M. Malekjani, J. Sol\`{a} Peracaula, Phys. Rev. D \textbf{100}, 023539 (2019).



\bibitem{Prigogine_1988-1989}  
I. Prigogine, J. Geheniau, E. Gunzig, P. Nardone, Proc. Natl. Acad. Sci. U.S.A. \textbf{85}, 7428 (1988).
\bibitem{Lima1992-1996}   
M. O. Calv\~{a}o, J. A. S. Lima, I. Waga, Phys. Lett. A \textbf{162}, 223 (1992);
J. A. S. Lima, A. S. M. Germano, L. R. W. Abramo, Phys. Rev. D \textbf{53}, 4287 (1996).
%
\bibitem{LimaOthers2023}   
T. Harko, Phys. Rev. D \textbf{90}, 044067 (2014); 
S. R. G. Trevisani, J. A. S. Lima, Eur. Phys. J. C \textbf{83}, 244 (2023). 
%
\bibitem{Freaza2002Cardenas2020}  
M. P. Freaza, R. S. de Souza, I. Waga, Phys. Rev. D \textbf{66}, 103502 (2002); 
V. H. C\'{a}rdenas, M. Cruz, S. Lepe, S. Nojiri, S. D. Odintsov, Rev. D \textbf{101}, 083530 (2020).
%




\bibitem{BarrowLima}  
J. D. Barrow, Phys. Lett. B  \textbf{180}, 335 (1986);
J. A. S. Lima, R. Portugal, I. Waga,  Phys. Rev. D \textbf{37}, 2755 (1988).
%
\bibitem{BrevikNojiri} 
I. Brevik, S. D. Odintsov,  Phys. Rev. D \textbf{65}, 067302 (2002);
S. D. Odintsov, D. S\'{a}ez-Chill\'{o}n G\'{o}mez, G. S. Sharov, Phys. Rev. D \textbf{101}, 044010 (2020).
%
\bibitem{EPJC2022} 
J. Yang, Rui-Hui Lin, Xiang-Hua Zhai, Eur. Phys. J. C \textbf{82}, 1039 (2022). 






\bibitem{EassonCai}  
D. A. Easson, P. H. Frampton, G. F. Smoot, Phys. Lett. B \textbf{696}, 273 (2011); 
Y. F. Cai, E. N. Saridakis,  Phys. Lett. B \textbf{697}, 280 (2011).
\bibitem{Basilakos1} 
S. Basilakos, D. Polarski, J. Sol\`{a}, Phys. Rev. D \textbf{86}, 043010 (2012);
S. Basilakos, J. Sol\`{a}, Phys. Rev. D \textbf{90}, 023008 (2014).




\bibitem{Koma45}  N. Komatsu, S. Kimura, Phys. Rev. D \textbf{87}, 043531 (2013), Phys. Rev. D \textbf{88}, 083534 (2013). 
\bibitem{Koma6}  N. Komatsu, S. Kimura, Phys. Rev. D \textbf{89}, 123501 (2014). 
\bibitem{Koma7}  N. Komatsu, S. Kimura, Phys. Rev. D \textbf{90}, 123516 (2014).
\bibitem{Koma8}  N. Komatsu, S. Kimura, Phys. Rev. D \textbf{92}, 043507 (2015).
\bibitem{Koma9}  N. Komatsu, S. Kimura, Phys. Rev. D \textbf{93}, 043530 (2016).



\bibitem{Neto2022}
E. M. C. Abreu, J. A. Neto, Phys. Lett. B \textbf{824}, 136803 (2022).
%
\bibitem{Gohar2024}  
H. Gohar, V. Salzano, Phys. Rev. D \textbf{109}, 084075 (2024).
%




%
\bibitem{Cai2005}
R. G. Cai, S. P. Kim, J. High Energy Phys. 02 (2005) 050.
%
\bibitem{Cai2011}
R. G. Cai, L. M. Cao, N. Ohta, Phys. Rev. D \textbf{81}, 061501(R) (2010).
%
\bibitem{Dynamical-T-2007}
R. G. Cai, L. M. Cao, Phys. Rev. D \textbf{75}, 064008 (2007).

\bibitem{Dynamical-T-20092014}
R. G. Cai, L. M. Cao, Y. P. Hu, Classical Quantum Gravity \textbf{26}, 155018 (2009);
S. Mitra, S. Saha, S. Chakraborty, Phys. Lett. B \textbf{734}, 173 (2014).
%
%
\bibitem{Sheykhi1}
A. Sheykhi, Phys. Rev. D \textbf{81}, 104011 (2010);
S. Mitra, S. Saha, S. Chakraborty, Mod. Phys. Lett. A \textbf{30}, 1550058 (2015).
\bibitem{Sheykhi2Karami} 
K. Karami, A. Abdolmaleki, Z. Safari, S. Ghaffari, J. High Energy Phys. 08 (2011) 150.
A. Sheykhi, S. H. Hendi, Phys. Rev. D \textbf{84}, 044023 (2011).
%
%
\bibitem{Santos2022}
P. S. Ens, A. F. Santos, Phys. Lett. B \textbf{835},137562 (2022).
%
%
\bibitem{Sheykhia2018} 
A. Sheykhi, Phys. Rev. D \textbf{103}, 123503 (2021); Phys. Rev. D \textbf{107},  023505 (2023);
A. Sheykhi, Phys. Lett. B \textbf{785}, 118 (2018); \textbf{850}, 138495 (2024).
%
\bibitem{ApparentHorizon2022} 
S. Nojiri, S. D. Odintsov, T. Paul, Phys. Lett. B \textbf{835}, 137553 (2022).
%

\bibitem{Cai2007} 
M. Akbar, R. G. Cai, Phys. Rev. D \textbf{75} 084003 (2007).
%
\bibitem{Cai2007B} 
M. Akbar, R. G. Cai, Phys. Lett. B \textbf{648}, 243 (2007).
%
\bibitem{Cai2008}
R. G. Cai, L. M. Cao, Y. P. Hu, J. High Energy Phys. 08 (2008) 090.
%
\bibitem{Sanchez2023}
L. M. S\'{a}nchez, H. Quevedo, Phys. Lett. B \textbf{839}, 137778 (2023).
\bibitem{Nojiri2024}
S. Nojiri, S. D. Odintsov, T. Paul, S. SenGupta, Phys. Rev. D \textbf{109}, 043532 (2024).
\bibitem{Odintsov2023ab}
S. D. Odintsov, D'Onofrio, T. Paul, Physics of the Dark Universe \textbf{42} (2023) 101277.
\bibitem{Odintsov2024}
S. D. Odintsov, T. Paul, S. SenGupta, Phys. Rev. D \textbf{109}, 103515 (2024).
\bibitem{Odintsov2024B}
S. D. Odintsov, S. D'Onofrio, T. Paul, Phys. Rev. D \textbf{110}, 043539 (2024).


\bibitem{Mohammadi2023}
A. Khodam-Mohammadi, M. Monshizadeh, Phys. Lett. B \textbf{843},138066 (2023).
%



\bibitem{Easther1-Egan1} 
R. Easther, D. Lowe, Phys. Rev. Lett. \textbf{82}, 4967 (1999); 
C. A. Egan, C. H. Lineweaver, Astrophys. J. \textbf{710}, 1825 (2010).
\bibitem{Pavon2013Mimoso2013}
J. P. Mimoso, D. Pav\'{o}n, Phys. Rev. D \textbf{87}, 047302 (2013).
\bibitem{Bamba2018Pavon2019}  
K. Bamba, A. Jawad, S. Rafique, H. Moradpour, Eur. Phys. J. C \textbf{78}, 986 (2018);
M. Gonzalez-Espinoza, D. Pav\'{o}n, Mon. Not. R. Astron. Soc. \textbf{484}, 2924 (2019).
\bibitem{deSitter_entropy}  
L. Dyson, M. Kleban, L. Susskind, J. High Energy Phys. 10 (2002) 011.
%
\bibitem{Saridakis2019}
S. Pan, W. Yang, C. Singha, E. N. Saridakis, Phys. Rev. D \textbf{100}, 083539 (2019).
\bibitem{Saridakis2021}
E. N. Saridakis, S. Basilakos, Eur. Phys. J. C \textbf{81}, 644 (2021).
%
\bibitem{Sharif2024}
M. Sharif, M. Zeeshan Gul, N. Fatima, Eur. Phys. J. C \textbf{84},1065 (2024).






%
\bibitem{Padma2010}    
T. Padmanabhan, Mod. Phys. Lett. A \textbf{25}, 1129 (2010).
\bibitem{Verlinde1} 
E. Verlinde, J. High Energy Phys. 04 (2011) 029.


\bibitem{HDE}  
A. Sayahian Jahromi, S. A. Moosavi, H. Moradpour, J. P. Morais Gra\c{c}a, I. P. Lobo, I. G. Salako, A. Jawad, Phys. Lett. B \textbf{780}, 21 (2018).
%
\bibitem{Padmanabhan2004}  
T. Padmanabhan, Classical Quantum Gravity \textbf{21}, 4485 (2004).
\bibitem{ShuGong2011}  
Fu-Wen Shu, Y. Gong, Int. J. Mod. Phys. D \textbf{20}, 553 (2011).


\bibitem{Koma14}  N. Komatsu, Phys. Rev. D \textbf{100}, 123545 (2019).
\bibitem{Koma15}  N. Komatsu, Phys. Rev. D \textbf{102}, 063512 (2020).
\bibitem{Koma16}  N. Komatsu, Phys. Rev. D \textbf{103}, 023534 (2021).
\bibitem{Koma17}  N. Komatsu, Phys. Rev. D \textbf{105}, 043534 (2022).
\bibitem{Koma19}  N. Komatsu, Phys. Rev. D \textbf{108}, 083515 (2023).  
\bibitem{Koma20}  N. Komatsu, Phys. Rev. D \textbf{109}, 023505 (2024). 

\bibitem{Padma2012AB}  T. Padmanabhan, arXiv:1206.4916 [hep-th]; Res. Astron. Astrophys. \textbf{12}, 891 (2012).

\bibitem{Cai2012}                                                                               
R. G. Cai, J. High Energy Phys. 1211 (2012) 016; 
S. Chakraborty, T. Padmanabhan, Phys. Rev. D \textbf{92}, 104011 (2015).
\bibitem{Hashemi}  
M. Hashemi, S. Jalalzadeh, S. V. Farahani, Gen. Relativ. Gravit. \textbf{47}, 53 (2015).
\bibitem{Moradpour}            
H. Moradpour, Int. J. Theor. Phys. \textbf{55}, 4176 (2016). 
\bibitem{Wang} 
J. Wang, Gen. Relativ. Gravit. \textbf{49}, 145 (2017).


\bibitem{Koma10}  N. Komatsu, Eur. Phys. J. C \textbf{77}, 229 (2017).
\bibitem{Koma11}  N. Komatsu, Phys. Rev. D \textbf{96}, 103507 (2017).
\bibitem{Koma12}  N. Komatsu, Phys. Rev. D \textbf{99}, 043523 (2019).
%
\bibitem{Koma18}  N. Komatsu, Eur. Phys. J. C \textbf{83}, 690 (2023).


\bibitem{Krishna20172019} 
P. B. Krishna, T. K. Mathew, Phys. Rev. D \textbf{96}, 063513 (2017); Phys. Rev. D \textbf{99}, 023535 (2019).
%
%
\bibitem{Mathew2022}
P. B. Krishna, V. T. H. Basari, T. K. Mathew, Gen. Relativ. Gravitat. \textbf{54}, 58 (2022).
\bibitem{Chen2022}
G. R. Chen, Eur. Phys. J. C \textbf{82}, 532 (2022).
\bibitem{Luciano}
G. G. Luciano, Phys. Lett. B \textbf{838}, 137721 (2023).
%
\bibitem{Mathew2023}
M. Muhsinath, V. T. H. Basari, T. K. Mathew, Gen. Relativ. Gravitat. \textbf{55}, 43 (2023).
%
\bibitem{Mathew2023b}   
M. T. Manoharan, N. Shaji, T. K. Mathew, Eur. Phys. J. C \textbf{83}, 19 (2023).


\bibitem{Pad2017}
T. Padmanabhan, Comptes Rendus Physique \textbf{18}, 275 (2017).
\bibitem{Tu2018}
Fei-Quan Tu, Yi-Xin Chen, Bin Sun, You-Chang Yang, Phys. Lett. B \textbf{784}, 411 (2018).
\bibitem{Tu2019}
Fei-Quan Tu, Yi-Xin Chen, Qi-Hong Huang, Entropy \textbf{21}, 167 (2019).


\bibitem{Chen2024}   
J. Chen, G. Chen, Eur. Phys. J. C \textbf{84}, 1178 (2024).







\bibitem{Hawking1Bekenstein1}  
S. W. Hawking, Phys. Rev. Lett. \textbf{26}, 1344 (1971); Nature (London) \textbf{248}, 30 (1974); Commun. Math. Phys. \textbf{43}, 199 (1975); Phys. Rev. D \textbf{13}, 191 (1976); 
J. D. Bekenstein, Phys. Rev. D \textbf{7}, 2333 (1973); Phys. Rev. D \textbf{9}, 3292 (1974);  Phys. Rev. D \textbf{12}, 3077 (1975).



\bibitem{Hooft-Bousso}
G. 't Hooft, Conf. Proc. C \textbf{930308}, 284 (1993) [arXiv:gr-qc/9310026]; L. Susskind, J. Math. Phys. \textbf{36}, 6377 (1995); R. Bousso, Rev. Mod. Phys. \textbf{74}, 825 (2002).



\bibitem{GibbonsHawking1977} 
G. W. Gibbons, S. W. Hawking, Phys. Rev. D \textbf{15}, 2738 (1977).



\bibitem{Dynamical-T-1998}
S. A. Hayward, Classical Quantum Gravity \textbf{15}, 3147, (1998).
%
\bibitem{Dynamical-T-2008}
S. A. Hayward, R. D. Criscienzo, M. Nadalini, L. Vanzo, S. Zerbini, Classical Quantum Gravity \textbf{26}, 062001 (2009).
%







\bibitem{Ryden1}  B. Ryden, \textit{Introduction to Cosmology} (Addison-Wesley, Reading, MA, 2002). 



\bibitem{Barrow22}  J. D. Barrow, T. Clifton, Phys. Rev. D \textbf{73}, 103520 (2006). 
\bibitem{Wang0102}  B. Wang, Y. Gong, E. Abdalla, Phys. Lett. B  \textbf{624}, 141 (2005).
\bibitem{Dynamical20052013} 
J. A. S. Lima, S. Basilakos, J. Sol\`{a},  Mon. Not. R. Astron. Soc. \textbf{431}, 923 (2013).







%
\bibitem{Das2008} S. Das, S. Shankaranarayanan, S. Sur, Phys. Rev. D \textbf{77}, 064013 (2008).
\bibitem{Radicella2010} N. Radicella, D. Pav\'{o}n, Phys. Lett. B \textbf{691}, 121 (2010).
%
\bibitem{LQG2004_123} 
A. Chatterjee, P. Majumdar, Phys. Rev. Lett. \textbf{92}, 141301 (2004).
%
\bibitem{Tsallis2012}  C. Tsallis, L. J. L. Cirto, Eur. Phys. J. C \textbf{73}, 2487 (2013). 
\bibitem{Czinner1Czinner2}       
T. S. Bir\'{o}, V. G. Czinner, Phys. Lett. B \textbf{726}, 861 (2013).
\bibitem{Barrow2020} J. D. Barrow, Phys. Lett. B \textbf{808}, 135643 (2020).

\bibitem{Nojiri2022}
S. Nojiri, S. D. Odintsov, V. Faraoni, Phys. Rev. D \textbf{105}, 044042 (2022).
\bibitem{Gohar2023}   
I. \c{C}imdiker, M. P. D\c{a}browski, H. Gohar, Eur. Phys. J. C \textbf{83}, 169 (2023).





\bibitem{Callen} H. B. Callen, \textit{Thermodynamics and an introduction to thermostatistics}, 2nd ed. (Wiley, New York, 1985).





\bibitem{GibbonsHawking1977Action} 
G. W. Gibbons, S. W. Hawking, Phys. Rev. D \textbf{15}, 2752 (1977).
%
\bibitem{York1986} 
J. W. York, Phys. Rev. D \textbf{33}, 2092 (1986).
\bibitem{York1988} 
B. F. Whiting, J. W. York, Phys. Rev. Lett. \textbf{61}, 1336 (1988).
\bibitem{Whiting1990} 
B. F. Whiting, Class. Quantum Grav. \textbf{7}, 15 (1990).
\bibitem{Wei2022} 
Shao-Wen Wei, Yu-Xiao Liu, R. B. Mann, Phys. Rev. Lett. \textbf{129}, 191101 (2022).



\bibitem{Broglie} L. de Broglie, \textit{Thermodynamics of Isolated Particle (Hidden Thermodynamics of Particles)}, (Gauthier-Villars, Paris, 1964).
%
\bibitem{ActionEntropy} 
D. Acosta, P. F. de Cordoba, J. M. Isidro, J. L. G. Santander, Int. J. Geom. Meth. Mod. Phys. \textbf{10}, 1350007 (2013).





\bibitem{RyuTakayanagi2006}
S. Ryu, T. Takayanagi, Phys. Rev. Lett. \textbf{96}, 181602 (2006);
J. High Energy Phys. 08 (2006) 045.


\bibitem{Arias2020} 
C. Arias, F. Diaz, P. Sundell, Classical Quantum Gravity \textbf{37}, 015009 (2020).
%


\bibitem{Relative_entropy}
D. D. Blanco, H. Casini, Ling-Yan Hung, R. C. Myers, J. High Energy Phys. 08 (2013) 060; 
D. L. Jaeris, A. Lewkowycz, J. Maldacena, S. J. Suh, J. High Energy Phys. 06 (2016) 004.







\end{thebibliography}
\end{document}